\documentclass[]{jfm}
\usepackage{float,pdflscape}
\usepackage{graphicx}
\usepackage[percent]{overpic}
\usepackage{newtxtext}
\usepackage{newtxmath}
\usepackage{natbib}
\usepackage{float}
\usepackage{hyperref}
\hypersetup{
    colorlinks = true,
    urlcolor   = red,
    citecolor  = cyan,
    linkcolor  = red,
}

\newcommand{\RomanNumeralCaps}[1]
\linenumbers
\usepackage{epstopdf}
\AppendGraphicsExtensions{.tiff}
\AppendGraphicsExtensions{.tif}
\usepackage{mathrsfs} 
\usepackage[dvipsnames]{xcolor}
\usepackage{subfigure}
\usepackage{float}
\usepackage{tabularx}
\usepackage{multicol}
\usepackage{multirow}
\usepackage{textcomp}
\makeatletter
\newcommand*{\rom}[1]{\expandafter\@slowromancap\romannumeral #1@}
\makeatother
\usepackage{setspace}
 \normalsize
\usepackage{booktabs}
\usepackage{multirow}
\usepackage {lineno}

\usepackage{caption}
\usepackage{lscape}

\usepackage{titlesec}  
\titlespacing{\chapter}{0pt}{10pt plus 0pt minus 0pt}{2pt plus 0pt minus 0pt}
\titlespacing{\section}{0pt}{8pt plus 0pt minus 0pt}{2pt plus 0pt minus 0pt}

\titlespacing{\subsection}{0pt}{6pt plus 0pt minus 0pt}{2pt plus 0pt minus 2pt}
\usepackage{amsmath,bm}
\expandafter\def\expandafter\normalsize\expandafter{%
  \normalsize  
  \setlength\abovedisplayskip{1pt}
 \setlength\belowdisplayskip{1pt}
  \setlength\abovedisplayshortskip{0pt}
  \setlength\belowdisplayshortskip{0pt}
}
\numberwithin{equation}{section}

\usepackage{fontawesome5}
\usepackage{tikz}
\usetikzlibrary{decorations.pathreplacing,calligraphy,backgrounds}

\title{New insights into the cavitation erosion by bubble collapse at moderate stand-off distances}

    \author{Zhesheng Zhao\aff{1}\thanks{These authors contributed equally to this work as co-first authors}, Shuai Li\aff{1}\footnotemark[1], Chengwang Xiong\aff{1,2}\corresp{\email{chengwang.xiong@hrbeu.edu.cn}}, Pu Cui\aff{1}, Shiping Wang\aff{1,2}, A-Man Zhang\aff{1,2}}

\affiliation{
\aff{1}College of Shipbuilding Engineering, Harbin Engineering University, 150001 Harbin, PR China\\
\aff{2}Nanhai Institute of Harbin Engineering University, Sanya 572024, PR China\\
}

\begin{document}

\maketitle

\begin{abstract}
Non-spherical bubble collapses near solid boundaries, generating water hammer pressures and shock waves, were recognized as key mechanisms for cavitation erosion. 
However, there is no agreement on local erosion patterns, and cavitation erosion damage lacks quantitative analysis.  
In our experiments, five distinct local erosion patterns were identified on aluminum sample surfaces, resulting from the collapse of laser-induced cavitation bubbles at moderate stand-off distances of $0.4\le\gamma\le2.2$, namely Bipolar, Monopolar, Annular, Solar-Halo, and Central. 
Among them, the Bipolar and Monopolar patterns exhibit the most severe cavitation erosion when the toroidal bubbles undergo asymmetrical collapse along the circumferential direction during the second cycle.
Shadowgraphy visualization revealed that asymmetrical collapse caused shockwave focusing through head-on collision and oblique superposition of wavefronts.  
This led to the variations in toroidal bubble radii and the positions of maximum erosion depth not matching at certain stand-off distances. 
Both initial plasma asymmetry and bubble-wall stand-off distance were critical in determining circumferential asymmetrical collapse behaviors.
At large initial aspect ratios, the elliptical jet tips form during the contraction process, resulting in the toroidal bubble collapsing from regions with smaller curvature radii, ultimately converging to the colliding point along the circumferential direction. 
Our three-dimensional simulations using OpenFOAM successfully reproduce the key features of circumferentially asymmetrical bubble collapse. 
This study provides new insights into the non-spherical near-wall bubble collapse dynamics and provides a foundation for developing predictive models for cavitation erosion.

\end{abstract}


\section{Introduction}\label{sec:introduction} %%%%%%%%%%引言
Cavitation refers to the phase change phenomenon that occurs when the local pressure of a liquid falls below a critical value at a specific temperature and flow velocity.
Over a century ago, \cite{silberrad1912propeller} reported on the erosion of propellers. 
\cite{rayleigh1917viii} identified bubble collapse as the primary cause of corrosion in hydraulic machinery such as hydrofoils and propellers. 
He derived an influential equation, known as the Rayleigh equation, for predicting the ideal collapse of spherical bubbles. This equation is based on the assumption of incompressible and inviscid flow.
Subsequently, various improved bubble pulsation equations were proposed, incorporating the effects of fluid viscosity, surface tension, compressibility, and complex boundary conditions \citep{gilmore1952growth, keller1980bubble, prosperetti1986bubble, zhang2024theoretical, han2024theoretical}, significantly enhancing the accuracy of bubble dynamics modeling. 

\begin{figure}
    \centering
    \includegraphics[width=1\linewidth]{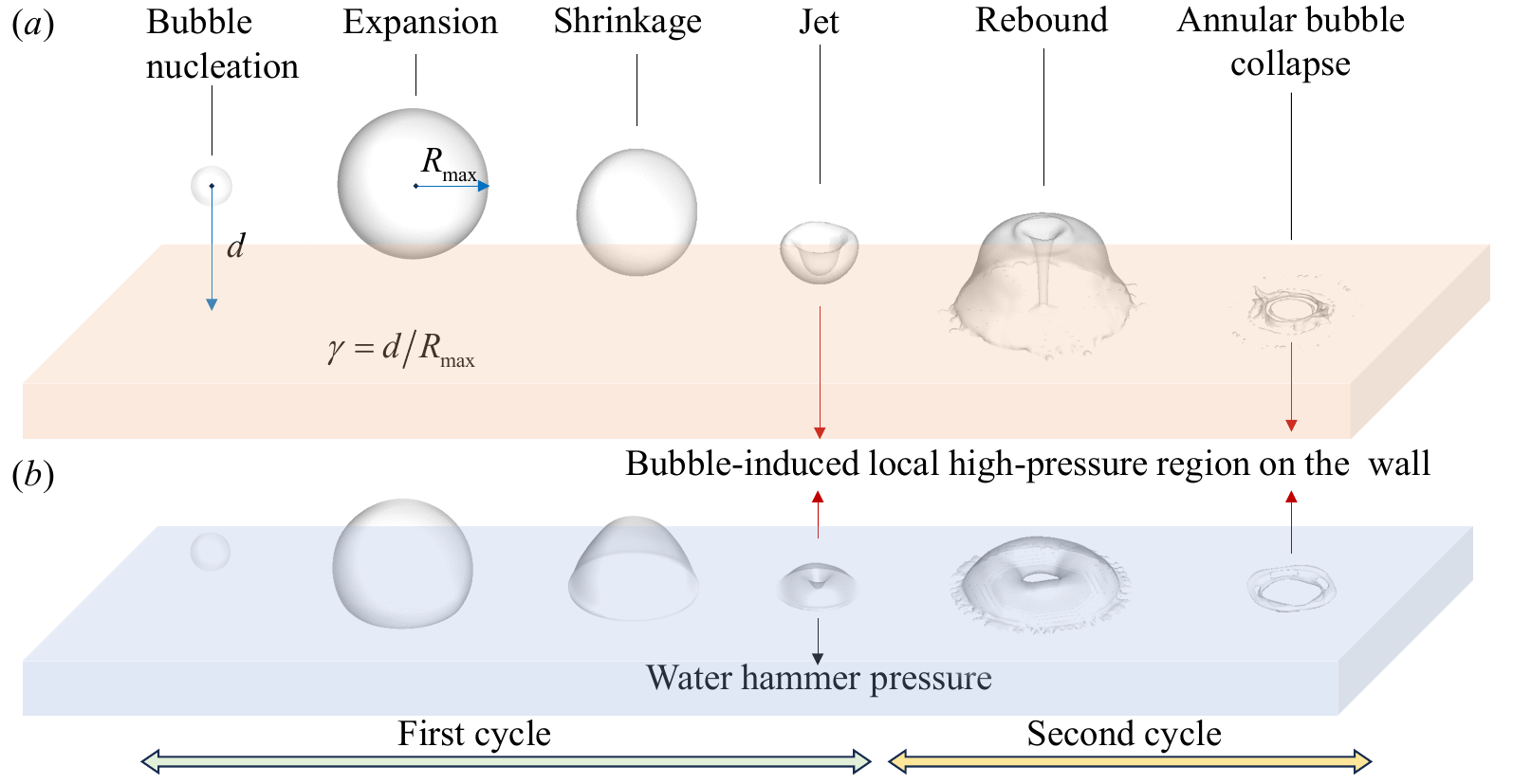}
    \caption{Schematic diagram of bubble pulsation and impulsive loads to induce cavitation erosion near the solid wall with the non-dimensional bubble-wall stand-off distance at ($a$) $\gamma>1$  and ($b$) $\gamma<1$.}
    \label{fig:Cavitation erosion diagram}
\end{figure}

A cavitation bubble undergoes non-spherical oscillations when it nucleates in the proximity of a solid boundary, as schematically illustrated in figure \ref{fig:Cavitation erosion diagram}. 
Upon the formation of a high-pressure bubble, it rapidly expands to its maximum volume due to the pressure difference between the bubble's gas and the ambient flow (frame 2). The dimensionless stand-off distance parameter, $\gamma$, is defined as the ratio of the distance $d$ (from the bubble nucleation site to the wall) to the maximum bubble radius $R_{\max}$. 
Subsequently, the bubble contracts under the reversed pressure difference. 
During the bubble contraction process, the pressure above the bubble becomes higher than the pressure below. This pressure difference drives the bubble's center of mass to migrate towards the wall under the influence of the Bjerknes force \citep{blake1986transient, tomita2002growth}. 
At the same time, as the bubble deviates from a spherical shape, the curvature at its top increases. Particularly when $\gamma < 1$, the bubble deforms into a conical shape, as shown in figure \ref{fig:Cavitation erosion diagram}($b$). This accelerates the bubble's collapse rate and induces significant flow field disturbances \citep{lauterborn1982cavitation, lechner2020jet}, facilitating the formation of a liquid jet with velocities reaching several hundred meters per second \citep{li2016analysis, lechner2019fast}, as illustrated in frames 3-4. 
For $\gamma > 1$ (figure \ref{fig:Cavitation erosion diagram}$a$), the jet impacts the lower surface of the bubble during the collapse process \citep{yin2021experimental}, generating a strong shock wave. However, the bubble still maintains a certain distance from the wall. 
During the subsequent expansion, the jet reaches the wall but with significantly reduced velocity due to the resisting water layer. 
In contrast, when $\gamma < 1$, the jet forms before the bubble reaches its minimum volume, accelerating to impact the wall with higher water hammer pressure \citep{dular2019high}, as shown in frame 4. 
Afterward, a toroidal bubble attached to the wall is formed \citep{reuter2017flow, sagar2020dynamics}. The violent collapse of this toroidal bubble generates strong shock waves and may damage the wall due to shear effects \citep{zeng2022wall}.
Subsequently, the bubble rebounds again, entering the second oscillation cycle, during which the toroidal ring undergoes a secondary collapse, once again releasing intense shock waves \citep{chahine2015modelling, trummler2020near}. 
The unique transient dynamic characteristics of cavitation bubbles present significant potential applicability in areas such as ultrasonic cleaning \citep{ohl2006surface}, cavitation peening \citep{takata2016effect}, emulsification \citep{orthaber2020cavitation}, and targeted drug delivery \citep{yualfred2022ultrasound}, thereby further advancing the development of this field.

The earliest research conducted by \cite{kornfeld1944destructive} revealed that direct impacts of jets on walls can cause erosion of materials. 
This erosion is attributed to the unstable deformation phenomena of vibrations induced by collapsing bubbles.
Researchers such as \cite{naude1961mechanism}, \cite{tomita1986mechanisms} and \cite{zhang2016experimental} utilized electric spark bubbles along with high-speed photography techniques to observe the pulsation of a single bubble in proximity to the wall. 
Their observations indicated that the water hammer pressure resulting from jets during the non-spherical collapse process is strong enough to cause plastic deformation in soft metals. 
\cite{shutler1965photographic} proposed that the collapse of toroidal bubbles near a wall produces powerful shockwave loads, leading to erosion damage that manifests in an annular distribution. 
As a result, the primary mechanisms of bubble-induced material erosion, namely microjets and toroidal bubble collapse, became the hotspots of subsequent research on cavitation. 
\cite{blake1987cavitation} successfully used the boundary integral method (BIM) to simulate bubble growth and collapse adjacent to a rigid wall. Subsequent advancements made by researchers, such as \cite{wang1996strong}, \cite{zhang2015study}, \cite{wang2016local}, \cite{li2021comparison}, and \cite{liu2024pressure}, incorporated weak compressibility into BIM, enabling the method to simulate bubble tearing phenomena caused by jets.

The introduction of the laser-induced cavitation method, pioneered by \cite{lauterborn1975experimental}, has endowed experimental research in cavitation erosion with non-invasiveness, high accuracy, and strong repeatability, facilitating more systematic investigations. 
There has been a consensus among previous literature \citep{lauterborn1982cavitation, vogel1989optical, tomita1986mechanisms, isselin1998laser} that the ring-like distribution of pits on metal surfaces generated by laser-induced bubbles resulted from the collapse of microbubble clusters formed during the contraction of the toroidal bubble. 
Subsequent studies conducted by \cite{philipp1998cavitation}, \cite{dular2019high} and \cite{ abedini2023situ} have systematically investigated the dynamic behavior of laser-induced bubbles adjacent to walls and their effect on surface damage of flat metal samples. 
The research indicates that microjets significantly contribute to cavitation erosion, particularly when $\gamma$ is less than 0.7 \citep{philipp1998cavitation}. 
As the distance increases, the shock wave loads emitted from the localized collapse of the toroidal bubble become the predominant factor responsible for cavitation erosion. Related numerical studies, such as  \cite{johnsen2009numerical}, \cite{sagar2020dynamics}, \cite{tong2022characteristics} and \cite{park2022numerical}, have also investigated the bubble collapse and jetting, aiming to elucidate the relationship between the pressure loads and material erosion. 
Most numerical results suggest the peak magnitudes of the jet and shockwave loads are in the same order at moderate $\gamma$ values, but both are generally below the yield strength of the metal.

Recent advancements in experimental equipment have greatly refined our understanding of bubble dynamics in close proximity to a wall. 
The velocity of needle-like jets can exceed 850 m/s when bubbles nucleate at $\gamma<0.053$, as reported by \cite{reuter2021supersonic}.
A similar phenomenon has also been observed in another experimental study utilizing optical ray tracing \citep{koch2021theory} and numerical study employing finite-volume method (FVM) \citep{lechner2020jet}. 
\cite{reuter2022cavitation} discovered that under specific conditions at $\gamma=0.06$, these needle jets can accelerate to velocities as high as 2400 m/s before impacting the wall. 
However, as pointed out by \cite{sieber2023cavitation}, the instability and atomization of these ultrahigh-speed jets weaken their impact pressure, and together with their inherently small momentum \citep{reuter2022cavitation}, make them ineffective in causing erosion. 
On the other hand, they have determined that the shock wave self-focusing mechanism, caused by the circumferential collapse of the toroidal bubble cluster at the end of the first cycle is the most significant factor in cavitation erosion.
This mechanism, revealed through sub-picosecond exposure shadowgraph imaging and hydrophone, was found to intensify the collapse of the residual gas phase and the emission of strong shock waves. 
\cite{dular2023bulk} suggested that when $ \gamma > 0.3$, cavitation erosion damage is also primarily caused by shock waves. Furthermore, beyond the yield strength, the acoustic impedance of different substrate materials influences the propagation and reflection of shock waves, thereby further affecting the severity of erosion.  
The most recent studies by \cite{y2023observing} and \cite{kuhlmann2024single} have focused on the
cavitation erosion in widely-used industrial materials like 316L stainless steel and nickel-aluminium bronze (NAB). 
Their results indicated that cavitation erosion at $\gamma < 0.3$ is primarily dominated by the first collapse, while the second collapse plays a more important role at $\gamma > 0.3$. 
As suggested by \cite{ganesh2016bubbly}, \cite{mihatsch2015cavitation} and \cite{ kuhlmann2023correlation}, although a single bubble creates only microscopic and scattered pits
on the material surface where intense local collapse occurs, these pits gradually accumulate over time, ultimately resulting in the fatigue deterioration of the material layer.

While the first collapse of a bubble extremely near the wall (with $\gamma < 0.3$) can cause intense localized loads and severe erosion, the high-temperature, high-pressure plasma generated during the initial bubble formation stage may also contribute to material damage \citep{reuter2022cavitation}. 
However, distinguishing between the erosion mechanisms at these two distinct stages is challenging. In practical cavitation scenarios, bubble growth is typically moderate and does not directly result in material damage. Damage often occurs after sheet cavitation is truncated by a re-entrant jet, followed by secondary shedding and complex vortex flow \citep{arabnejad2020hydrodynamic}, leading to the formation of microbubble clusters \citep{dular2015mechanisms, wang2023numerical, yang2025cavitation}. 
These microbubbles generally do not collapse very close to the wall during the first cycle but instead transmit collapse shock waves to the wall \citep{cao2017qualitative}. Furthermore, research by \cite{philipp1998cavitation} indicates that bubble collapse at larger stand-off distances ($\gamma > 0.3$) can still cause erosion as the bubble migrates toward the wall during the second cycle. Despite these insights, the mechanisms underlying the localized severe damage remain incompletely understood. Therefore, this study focuses on the dynamics of the second bubble collapse within the moderate stand-off distance range ($\gamma = 0.4 \sim 2.2$)  and its correlation with erosion characteristics.

%%当前存在的问题
Despite the considerable progress made in cavitation erosion research to date, many challenges remain, some of which are as follows: i) divergent understandings towards the roles of jet and shockwave in cavitation erosion; ii) inconsistent cavitation erosion patterns emerge from experiments conducted under identical stand-off distance parameters; iii) the conflict between the relatively low wall pressure loads obtained from numerical simulations and the obvious cavitation erosion observed from experiments at moderate $\gamma$; and iv) the lack of quantitative analyses on the local erosion distribution. 
Earlier studies proposed that the initial asymmetry nature of laser-induced bubbles primarily drives uneven collapse of the toroidal bubble, which has been confirmed in the present study through comprehensive experiments and numerical simulations. 
These results illustrate the dependency of typical local erosion characteristics on distance parameters and initial shape parameters, further deepening the understanding of the cavitation erosion formation mechanisms and providing new insights into the prediction and prevention of cavitation erosion.

%%%本文结构

The rest of the paper is organized as follows: \S\ref{sec:Methods} outlines the experimental and numerical methods utilized in this study; 
\S\ref{sec:Cavitation erosion mechanism} quantitatively analyzes the characteristics of cavitation erosion field distribution and identifies bubble collapse-induced shock waves as the primary cause of erosion; 
\S\ref{4} investigates the circumferentially asymmetric collapse mechanism of non-spherical initial bubbles and their induced wall loads; the conclusions are presented in \S\ref{sec:Conclusion}.

\section{Methodology}\label{sec:Methods} %%实验,数值

\subsection{Experimental set-up}\label{2.1} %%%实验方法介绍

\begin{figure}
    \centering
    \includegraphics[width=1\linewidth]{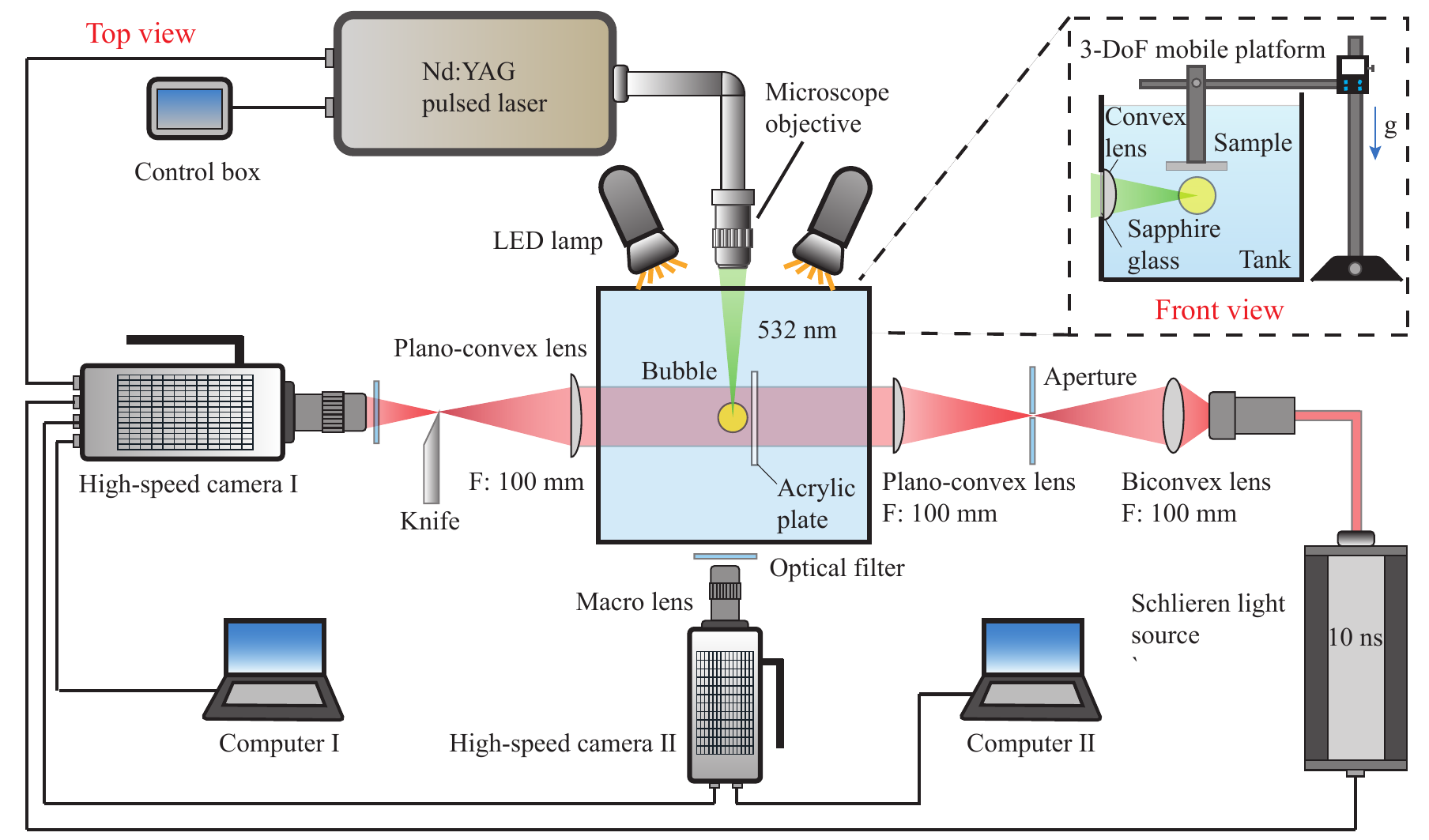}
    \caption{%The schematic diagram of experimental platform. 
    The schematic diagram of the experimental platform highlights several key components: a laser-focused cavitation bubble generation unit, a Schlieren imaging optical path, a dual-angle synchronized imaging system (plan-view), and a bubble-induced material erosion setup located in the upper right corner (side-view).}
    \label{fig: Platform diagram}
\end{figure}

Figure \ref{fig: Platform diagram} illustrates the experimental platform, which consists of a laser-focused unit for generating cavitation bubbles, a schlieren imaging system for capturing shock waves, a dual-angle synchronized imaging system, and a 3-DoF (degree of freedom) mobile platform for precisely adjusting the position of the sample plate.
To better visualize bubble collapse patterns and shock waves, a transparent acrylic plate was vertically mounted on the 3-DoF mobile platform, perpendicular to the horizontal plane.
Alternatively, the aluminum sample plate was mounted horizontally to study cavitation erosion, as shown in the subfigure in figure \ref{fig: Platform diagram}. 
These two scenarios are equivalent in terms of bubble dynamics at the corresponding $\gamma$ values since their optical path of the laser focusing is parallel to the plates and the buoyancy effect can be ignored at such small buoyancy parameter $\delta=\sqrt{\rho g R_{\max } / p_{\infty}}\sim \mathcal{O}(10^{-2})$ \citep{reuter2022cavitation}.
Experiments were conducted under ambient conditions, with temperatures ranging from $20^\circ$C to $25^\circ$C and a surrounding pressure of $p_\infty=10^5$ Pa.

A Nd:YAG pulsed laser (Nimma-900, pulse duration 8 ns) with a wavelength of 532 nm and pulse energy ranging from 20 to 40 mJ. The quartz water tank, measuring 100 mm × 100 mm × 100 mm, was filled with deionized water to a depth of 90 mm. 
To achieve a more precise laser focusing, a 30 mm diameter sapphire glass with a thickness of 0.25 mm was installed into a hole on the side of the quartz tank.
The laser beam passed through the microscope objective (M Plan Apo L 10 $\times$, NA = 0.28), the sapphire glass and a convex lens, was focused near the center of the water tank to generate cavitation bubbles with $R_{\max} \sim \mathcal{O}(1)$ mm. This laser-induced bubble system has been well validated in our previous work, as detailed in \cite{li2024cavitation} and \cite{zhang2025free}. 
Additionally, a number of convex lenses with different focal lengths ($\phi=25.4$ mm, $f$ = 30, 40 and 50 mm) were used in front of the sapphire glass to adjust the focusing angle of the laser, while the laser energy was continuously varied to alter the plasma shape.

A schlieren imaging system, designed to capture the shock waves emanating from bubble collapse, features an optical path depicted in figure \ref{fig: Platform diagram}. 
A nanosecond high-frequency pulsed laser (Cavilux HF, wavelength 810 nm, pulse duration 10 ns) was employed to enhance the sensitivity of the schlieren imaging instead of the traditional LED point light source. 
This laser is expanded by a plano-convex lens with a 100 mm focal length to create a parallel beam of approximately 20 mm in diameter that passes through the cavitation region.
The wavefront, possessing a density significantly higher than the ambient medium, results in non-uniformities in the refractive index of the surrounding flow field when the bubble collapses, thereby deflecting light rays. 
The light rays are subsequently focused by a schlieren lens (F = 100 mm) and intercepted by a sharp knife edge, producing distinctive dark streaks in the imagery.
The transient dynamics of bubble nucleation and collapse are captured by a high-speed camera \uppercase\expandafter{\romannumeral1} (Phantom V2012) coupled with a LAOWA macro lens (100 mm, F2.8), at a resolution of $128 \times 64 $ pixels, recording at 640,000 frames per second. 
Simultaneously, high-speed camera \uppercase\expandafter{\romannumeral2} (Phantom V12.1, frame rate 180,000 fps) was used to capture side-view bubble oscillations via the shadow method from another angle of view. 
Uniform background illumination across the water tank was provided by diffusing two 300W LED lights through frosted glass. 
The exposure of camera \uppercase\expandafter{\romannumeral2} was set to 1 $\upmu$s, and a filter was positioned in front of the lens to shield the sensor.

\begin{figure}
    \centering
    \includegraphics[width=1\linewidth]{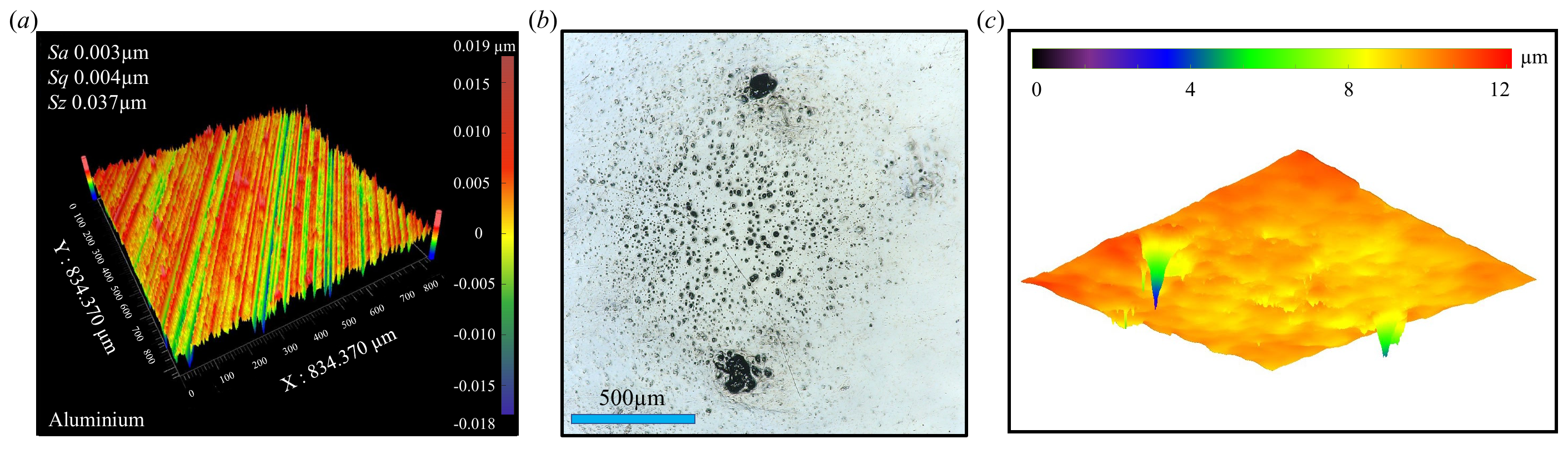}
    \caption{Surface characterization of a pure aluminium plate with ultra-smooth treatment: ($a$) the surface roughness before cavitation erosion; ($b$) the plane view of cavitation erosion distribution; ($c$) 3D reconstruction of cavitation erosion distribution. }
    \label{fig:Metal sample}
\end{figure}

The size of pure aluminium samples employed for cavitation erosion tests is 10 mm × 10 mm × 2 mm, and the surface is treated to achieve ultra-smoothness using a precision diamond machining tool (Moore Nanotech UP250L). 
Figure \ref{fig:Metal sample}(\textit{a}) manifests the surface roughness of the processed metal samples, measured by a white light interferometer, indicating a smooth, flat surface with nanometer precision and sub-micron geometric accuracy. 
To avoid contamination from water droplets and dust, the metal sample is cleaned with ethanol prior to the erosion tests. 
During the erosion experiment, side views were captured using camera \uppercase\expandafter{\romannumeral1} with the shadow method ($128 \times 128$ pixels, 430,000 fps), and the stand-off distance parameters were measured and adjusted.

Surface damage of the metal sample after erosion tests are examined using a super-depth-of-field microscope (Olympus DSX1000). 
Its measurement accuracy in the $x$-$y$ plane is $\pm 3\%$, with a repeatability of $\pm 2\%$, while the height measurement repeatability in the $z$-axis direction is $\sigma_{n-1} \leq 1 \ \upmu\text{m}$. 
As depicted in figure \ref{fig:Metal sample}(\textit{b}), the details of pitting holes are discernible. 
The microscope can also output high-resolution 3D images and depth contours through optimized imaging algorithms, as shown in figure \ref{fig:Metal sample}(\textit{c}), providing precise quantitative data for subsequent analysis.

\begin{table}
\centering
\setlength{\tabcolsep}{5mm}
\renewcommand{\arraystretch}{1.5}
\begin{tabular}{ccccc}
        
        & $p_{0}$ (Pa) & $B$ (MPa) & $\rho$ (kg/m$^3$) & $\gamma$ \\
        \hline
        Liquid phase & 1.01$\times 10^5$ & 305 & 998 & 7.15 \\
        Gas phase & 1.03$\times 10^4$ & 0 & 0.12 & 1.4 \\
        \hline
    \end{tabular}
    \caption{The coefficients for Tait state equation of liquid and gas phases.}
    \label{tab:parameters}
\end{table}

\subsection{Numerical methodology}\label{2.2} %%%%数值方法
The numerical simulation of the dynamic pulsation of cavitation bubbles in proximity to a solid boundary involves solving the compressible and viscous multiphase flow problem.  
The present study utilized \textsc{CavBubbleFoam} solver, which is a customized and rigorously validated version of \textsc{CompressibleInterFoam} built in OpenFOAM-4.X. 
Notably, this solver is specifically designed for bubble dynamics, incorporating modifications to the phase fraction field $\alpha$ and ensuring meticulous bubble mass conservation \citep{koch2016numerical, zeng2018wall,zeng2022wall, reese2022microscopic}.
Meanwhile, the solver accounts for surface tension but neglects heat and mass transfer. The governing equations comprise the continuity equation and the Navier-Stokes equations:
\begin{equation}
   \frac{\partial \rho}{\partial t}+\nabla \cdot(\rho \boldsymbol{u})=0
\label{continuity equation},
\end{equation}
\begin{equation}
   \frac{\partial \rho \boldsymbol{u}}{\partial t}+\nabla \cdot(\rho \boldsymbol{u} \boldsymbol{u})=-\nabla p+\nabla \cdot \boldsymbol{\tau}+\boldsymbol{f}_{\sigma},
\label{momentum equation}
\end{equation} 
where $\rho$ represents the fluid density, $\boldsymbol{u}$ denotes the velocity, $p$ is the pressure. 
$\boldsymbol{f}_{\sigma}$ represents the surface tension term, modeled using the continuous surface force (CSF) method \citep{brackbill1992continuum}, which can be expressed as:
\begin{equation}
   \boldsymbol{f}_{\sigma} = \sigma \left[ -\nabla \cdot \left( \frac{\nabla \alpha_1}{|\nabla \alpha_1|} \right) \right] \nabla \alpha_1,
\label{surface_tension}
\end{equation}
here $\sigma = 0.07$ N/m is the surface tension coefficient between the gas and water.
The viscous stress tensor $\boldsymbol{\tau}$ is given by:
\begin{equation}
    \boldsymbol{\tau} = \mu \left(\nabla \boldsymbol{u} + \nabla \boldsymbol{u}^{\mathrm{T}} - \frac{2}{3}(\nabla \cdot \boldsymbol{u}) \boldsymbol{I}\right),
    \label{momentum equation}
\end{equation} 
where $\mu$ is the dynamic viscosity and $\boldsymbol{I}$ is the identity tensor.
The Volume of Fluid (VoF) method is employed to capture and characterize the interface between gas and liquid phases: %\citep{weller2008new, miller2013pressure, zeng2022wall}: 
\begin{equation}
   \frac{\partial \alpha_{1}}{\partial t}+\nabla \cdot(\alpha_{1} \boldsymbol{u})+\nabla \cdot\left(\alpha_{1}\alpha_{2} \boldsymbol{U}_{r}\right)=\alpha_{1}\alpha_{2}\left(\frac{\psi_{1}}{\rho_{1}}-\frac{\psi_{2}}{\rho_{2}}\right) \frac{\mathrm{D} p}{\mathrm{D} t}+\alpha_{1} \nabla \cdot \boldsymbol{u}
\label{VOF},
\end{equation}
\begin{equation}
    \alpha_{2} = 1 - \alpha_{1},
    \label{VOF2}
\end{equation} 
where $\alpha_i$ represents the volume fraction of the phase, with subscripts $i=$1 and 2 denoting the liquid and gas phases, respectively, and 
$\boldsymbol{U}_{r}$ is the relative velocity between the two phases. 
The isothermal compressibility for both fluids, denoted as $\psi_i=\mathrm{D} \rho_i / \mathrm{D} p$, can be calculated using the Tait equation of state: 
\begin{equation}
   p=\left(p_{0}+B\right)\left(\frac{\rho}{\rho_{0}}\right)^{\gamma}-B.
\label{Tait equation}
\end{equation} 
The coefficients of equation (\ref{Tait equation}) are listed in table \ref{tab:parameters}.

\subsection{Numerical set-ups and validation} 
\label{2.3}

\begin{figure}
    \centering
    \includegraphics[width=0.93\linewidth]{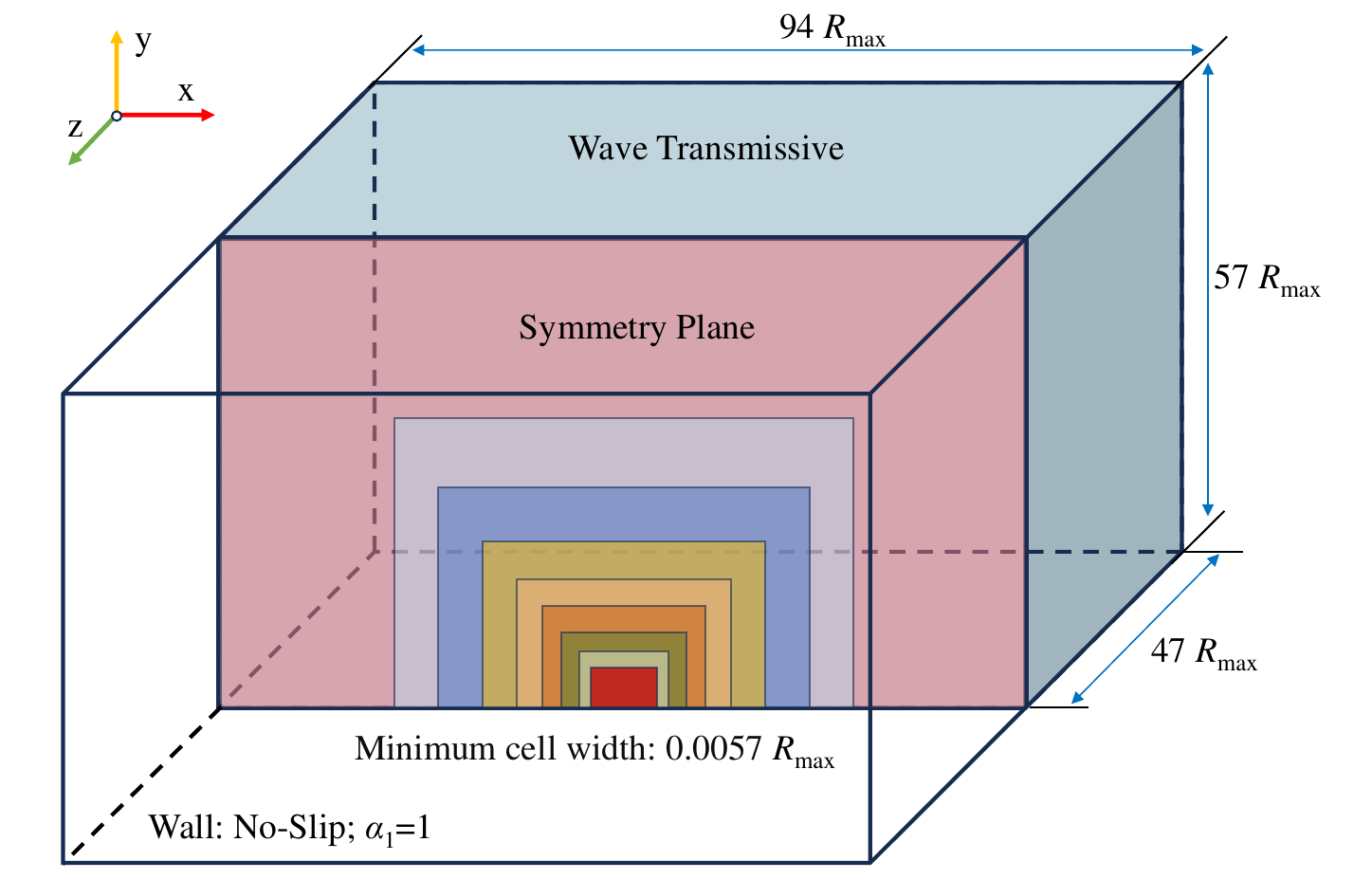}
    \caption{Schematic diagram of the numerical simulation domain. Levels of mesh refinement are indicated by the contours on the symmetry plane.}
    \label{fig: Mesh}
\end{figure}

A schematic diagram shown in figure \ref{fig: Mesh} illustrates the general set-ups for the numerical simulations.
The $x$-$o$-$y$ plane was designated as a symmetric boundary, allowing for sparing half the computational domain and improving computational efficiency.  
As a result, the size of the computational domain was set to $94 R_{\max} \times 57 R_{\max} \times 47 R_{\max}$, with the maximum equivalent radius of the simulated bubble being approximately 530 $\upmu$m in this study. 
Similar to \cite{beig2018temperatures}, wave-transmissive boundary conditions were implemented on all external boundaries, except the no-slip bottom wall, to emulate an extended fluid domain and minimize the impact of boundary constraints on the simulation outcomes. 
The simulation utilized a structured mesh with the coarsest cell with $\Delta x=0.73R_{\max}$ at the outermost boundary. 
After eight levels of successive refinement, the finest cell width in the vicinity of the bubble and the adjacent wall was reduced to $\Delta x=0.0057R_{\max}$, ensuring adequate resolution in these critical regions. 
In line with \cite{reuter2021supersonic} and \cite{reuter2019high}, a persistent liquid film, modeled by fixing $\alpha_1=1$ at the wall nodes, prohibits direct contact between the bubble and wall. 
In the simulation, the initial state of the laser-induced bubble was modeled by introducing a small volume of high-pressure, non-condensable gas into the flow field, with a phase fraction of $\alpha_{2} = 1$. 
Initially, we employed the spherical bubble theory \citep{keller1956damping, zhang2023unified} and iteratively adjusted the initial pressure $P_0$ and radius $R_0$ in an infinite domain. By setting $P_0 = 4800$ bar (within the value range used by \cite{lechner2020jet}, 
\cite{zeng2022wall} \& \cite{reese2022microscopic}
) and $R_0 = 0.055R_{\max}$, we achieved a good agreement between the simulated maximum radius and oscillation period with experimental data. In subsequent simulations, with the bubble's $R_{\max}$ measured at 530 \textmu m, we selected an initial radius $R_0=29$ \textmu m. 
The effect of phase transition, which contributes to vapor condensation in physical experiments, is accounted for—as demonstrated in the studies by \cite{raman2022microemulsification}, \cite{ohl2024cavitation}, and \cite{fan2024amplification}—by applying an attenuation coefficient of 0.34 to the bubble’s internal pressure at its maximum volume during the first and second expansions.

\begin{figure}
    \centering
    \includegraphics[width=0.95\linewidth]{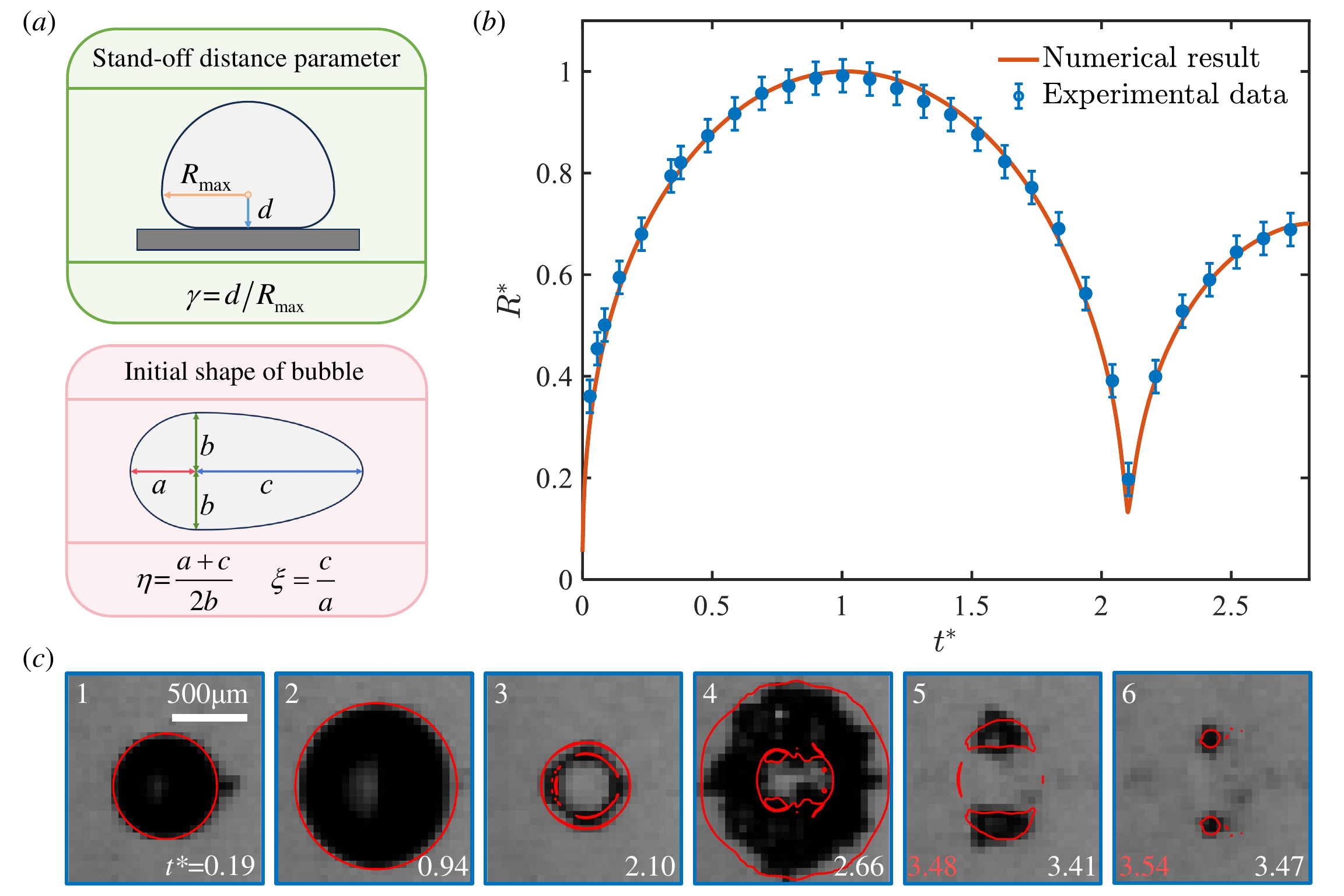}
    \caption{ $(a)$ Definitions of non-dimensional stand-off distance $\gamma$, 
    aspect ratio $\eta$ and the axial asymmetry ratio $\xi$. $(b)$ Comparison of the time evolution of the bubble radius obtained from the experiment and numerical simulation for the case of $\gamma=0.72$, $\eta=1.8$ and $\xi=1.9$. The error bar indicates the uncertainty in the experimental measurements. Nondimensional times, defined as $t^{*}=t/(R_{\max } \sqrt{\rho / P_{\infty}})$, and the dimensionless radius is expressed as $R^{*}=R/R_{\max}$. $(c)$ Comparison of experimental and numerical bubble profiles, where the red curve represents the numerical bubble iso-surface for $\alpha=0.5$, with experimental time in the bottom right and simulation time in red at the bottom left.
    }
    \label{fig: radius-}
\end{figure}

Figure \ref{fig: radius-}$(a)$ presents the non-dimensional parameters to characterize the cavitation bubble. 
It has been demonstrated in \cite{liang2022comprehensive} that the plasma generated by laser usually exhibits a non-spherical shape, resembling a spindle or teardrop shape.
In addition, the conventional stand-off distance, the aspect ratio $\eta = (a+c)/2b$ and the axial asymmetry parameter $\xi = {c}/{a}$ are introduced to more precisely reproduce the influence of the initial bubble geometry, which is also the focus of the present study.
The initial shape of a laser-induced cavitation bubble can be approximated by two conjoined semi-ellipsoids sharing an axis of length $2b$.
Specifically, $\eta$ reflects the overall flattening or longitudinal extension of the spindle shape, while $\xi$ quantifies the asymmetry between the left and right semi-ellipsoids. 
It should be noted that when dealing with small $\gamma$ values, the maximum bubble radius corresponds to the radius of the bubble's upper hemisphere at its maximal expansion, as shown in figure \ref{fig: radius-}$(a)$, \cite{reuter2022rayleigh} also employed this method, referring to it as a direct image measurement method.  
In experiments, the initial shape of the bubble is defined at the first frame after the plasma flashing.
To validate the accuracy of the numerical simulations, we selected conditions of $\gamma=0.72$, $\eta=1.8$, and $\xi=1.9$, and compared the results with experimental data. 
Figure \ref{fig: radius-}$(b)$ presents the time evolution of the bubble radius, demonstrating excellent agreement between numerical simulations and experimental data during the first oscillation cycle. Figure \ref{fig: radius-}$(c)$ further compares the bubble dynamics over the first two oscillation cycles, with the red contours representing the bubble interface from numerical simulations. 
During the expansion phase, the bubble appeared nearly circular in the top view (frames 1–2) and shrank to its minimum volume at $t^{*}=2.10$, forming a relatively uniform ring-shaped structure that matched well with the red contour from the experiment. In the second oscillation cycle, the initial asymmetry of the bubble was amplified, leading to increased curvature on both sides of the toroidal bubble (frame 4), which collapsed preferentially (frame 5), and finally converged circumferentially to two points perpendicular to the optical axis (frame 6). 
It is worth noting that the simulated bubble volume in the second oscillation cycle is slightly larger than the experimental measurements (frame 4), and the collapse speed is slightly lower, possibly due to the simplification of the bubble phase change process in numerical simulations.

\section{Characteristics of cavitation erosion and bubble collapse}\label{sec:Cavitation erosion mechanism}

\begin{figure}
    \centering
    \includegraphics[width=0.95\linewidth]{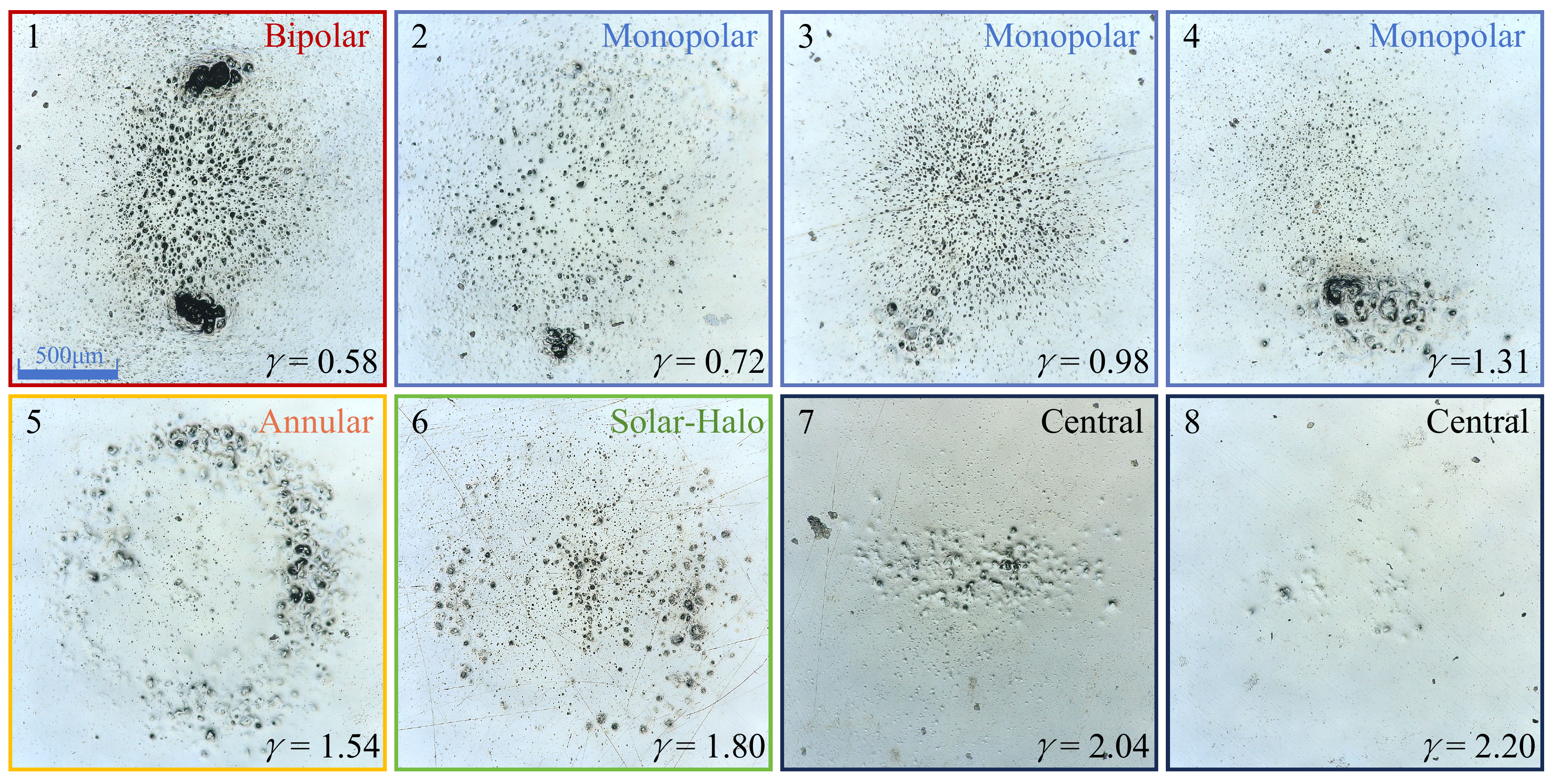}
    \caption{Typical distribution patterns of cavitation erosion on pure aluminium plate. Each sample underwent 900 times collapse of laser-induced bubble generated with the same energy. The laser is incident from left to right.}  
    \label{fig:cavitation erosion}
\end{figure}

\subsection{Cavitation erosion patterns}\label{3.1}

A series of cavitation erosion tests were conducted at 17 different stand-off distances ranging from $\gamma = 0.58$ to $2.20$, with an interval of approximately 0.1. In terms of the experimental findings presented in this section, the optical path was adjusted to maximize the sphericity of the laser-induced bubbles, aiming for values of both $\xi$ and $\eta$ approaching 1.
For each individual test, the laser was activated 900 times (repetition rate: 5 Hz) using the same energy level, resulting in an approximate maximum radius of the induced bubbles of 0.97 mm.
Based on microscopic observations of all tested specimens, as shown in figure \ref{fig:cavitation erosion}, the patterns of cavitation erosion were categorized into five typical types: Bipolar, Monopolar, Annular, Solar-Halo, and Central. 
Detailed description of the erosion patterns are elaborated as follows:

i) The Bipolar pattern is frequently observed at relatively low stand-off distances ($\gamma<0.8$) within the scope of the present study. It exhibits the most severe local erosion damage symmetrically aligned at the opposite sides of a circular erosion zone, whose connecting line is either parallel to or perpendicular to the axis of the laser beam. 
Within the circular cavitation zone, there are numerous densely distributed small pitting corrosion pits. However, these indentations have much smaller diameters and depths compared to the erosion in the ``polar'' regions.

ii) The Monopolar pattern occurs within a large stand-off parameter range ($\gamma=0.7\sim1.4$), with its primary characteristic being that the most severe erosion appears at one ``polar'' of the circular erosion zone.  
The distribution of small pits within the circular erosion area differs between the monopolar and bipolar patterns, with the former exhibiting less overall damage. 
The erosion distributions of the monopolar pattern at three different $\gamma$ are present in figure \ref{fig:cavitation erosion}, indicating significant differences in the extent and range of erosion in the ``polar'' region. 
Particularly, there is significant erosion at $\gamma=1.31$,  which is comparable to that observed in the Bipolar pattern.

iii) Annular erosion patterns dominate at $\gamma \approx 1.5$, with the erosion pits primarily distributed along a ``ring-like'' circumference.  
Inside the annular region, there are hardly any significant occurrences of erosion.
Compared to the Bipolar and Monopolar modes, the annular pattern generally possesses an axisymmetric feature, which is more in line with the conventional understanding of cavitation erosion.
However, the circumferential asymmetrical patterns, rather than the axisymmetric pattern, manifest the peak erosion damage.

iv) With the stand-off distance further increased to approximately $\gamma = 1.6\sim1.8$, the Solar-Halo pattern emerges, characterized by the concentrated distribution of pits at the center and along the circumference of the circular zone. 
However, no large or deep pits are observed in either of these areas.

v) For $\gamma$ exceeding 2.0, only small indentations appear at the center, which is referred to as the Center pattern.  At $\gamma$ = 2.2, the surface shows almost no significant damage, with only slight traces visible. 
It can be anticipated that increasing the stand-off distance further will make the cavitation erosion caused by bubble collapse negligible.

\begin{figure}
    \centering
    \includegraphics[width=1\linewidth]{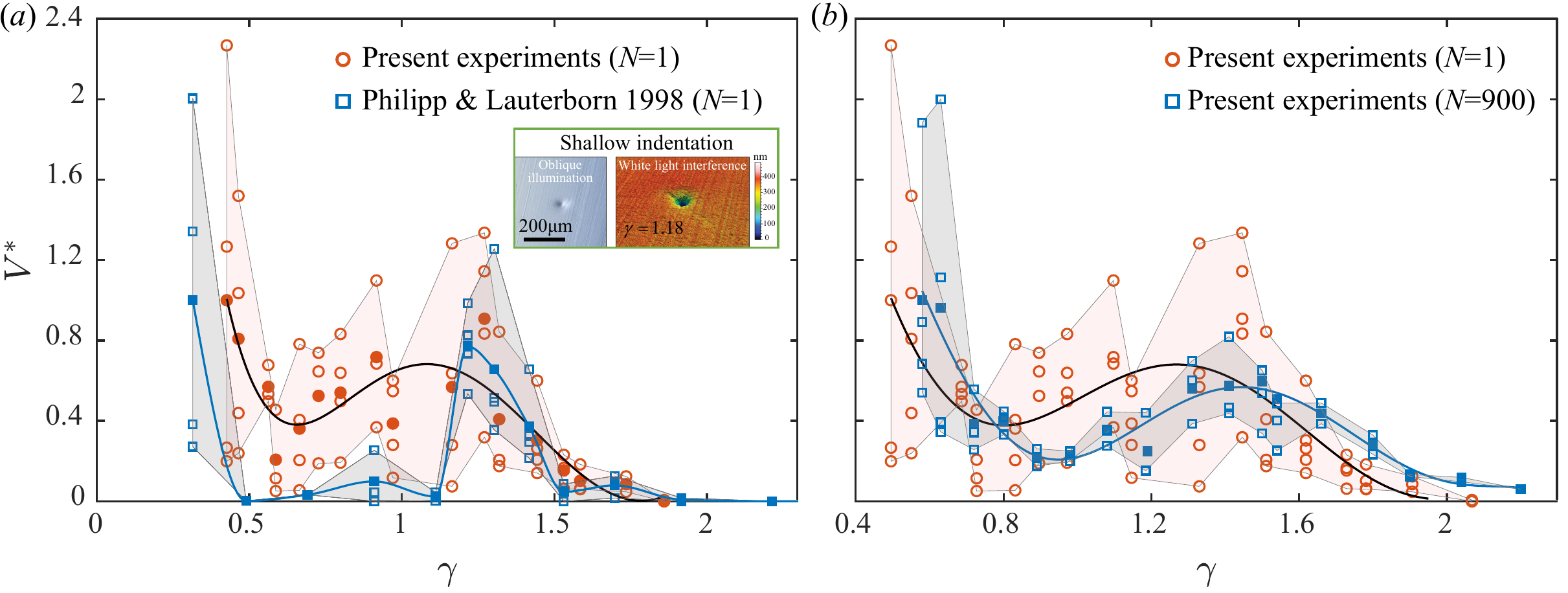}
    \caption{Variation of the overall volume of cavitation erosion pits within the range of moderate stand-off distances. Nondimensionalization is performed using the average damage volume corresponding to the smallest $\gamma$. 
    $(a)$ In comparison with the experimental results reported by \cite{philipp1998cavitation}, the bubble number is $N = 1$. The $R_{\max}$ is calculated using the formula: $R_{\max}=T_{\rm c}\sqrt{(p_\infty-p_v)/\rho}/0.915$. $(b)$ A comparison between $N = 1$ and $N = 900$ cases, where $R_{\max}$ is directly measured from the images. 
    The solid marks denote the corresponding average values, based on which the corresponding fitted curves are obtained.}
    \label{fig:Volume}
\end{figure}

\subsection{Quantification of cavitation erosion}\label{3-2}

In this subsection, quantitative analysis was performed on the cavitation erosion of the tested aluminium specimens. First, single-bubble induced cavitation erosion experiments were conducted. Since some shallow indentations have depths on the submicron scale, they are almost invisible under bright-field observation and can only be seen under oblique illumination. Therefore, we used a white light interferometer (gbs WLI) for 3D pit reconstruction and volume measurement. For comparison, the experimental results presented in figure \ref{fig:Volume}$(a)$ of \cite{philipp1998cavitation} have been reproduced in this study. 
It is worth noting that they employed a theoretical estimate of $R_{\max}$ in the free field, derived from the Rayleigh equation, to normalize the stand-off distance. 
Specifically, $R_{\max}=T_{\rm c}\sqrt{(p_\infty-p_v)/\rho}/0.915$, where $T_{\rm c}$ denotes the half-period of the first bubble oscillation cycle. 
In contrast, this study utilizes directly measured values of $R_{\max}$ for normalization. Since the Rayleigh equation tends to overestimate $R_{\max}$ for bubbles near a wall, the resulting normalized stand-off distance, defined as  $\gamma=d/R_{\max}$, is correspondingly underestimated.
To facilitate valid comparison, we also used the Rayleigh equation to calculate $R_{\max}$, and the average value of the data points at the corresponding smallest $\gamma$ to normalize the damage volume. 
The overall trend of the variation of erosion volume with respect to $\gamma$ observed in our study is consistent with that in \cite{philipp1998cavitation}. 
The maximum erosion volume occurs at the smallest $\gamma$ within the investigated range for both studies.
As $\gamma$ increases, the normalized cavitation erosion volume, $V^*$, initially decreases, then increases, and decreases again. Subsequently, another prominent peak appears at $\gamma \approx 1.3$, followed by a gradual decrease toward zero. 
However, there are some differences: the peak volume we measured at $\gamma = 0.5$ corresponds to the trough reported in \cite{philipp1998cavitation}. Additionally, within the range of $\gamma = 0.5$ to $1.3$, our results show a greater average value and higher dispersion, especially the peak around $\gamma = 0.9$.

\begin{figure}
    \centering
    \includegraphics[width=0.95\linewidth]{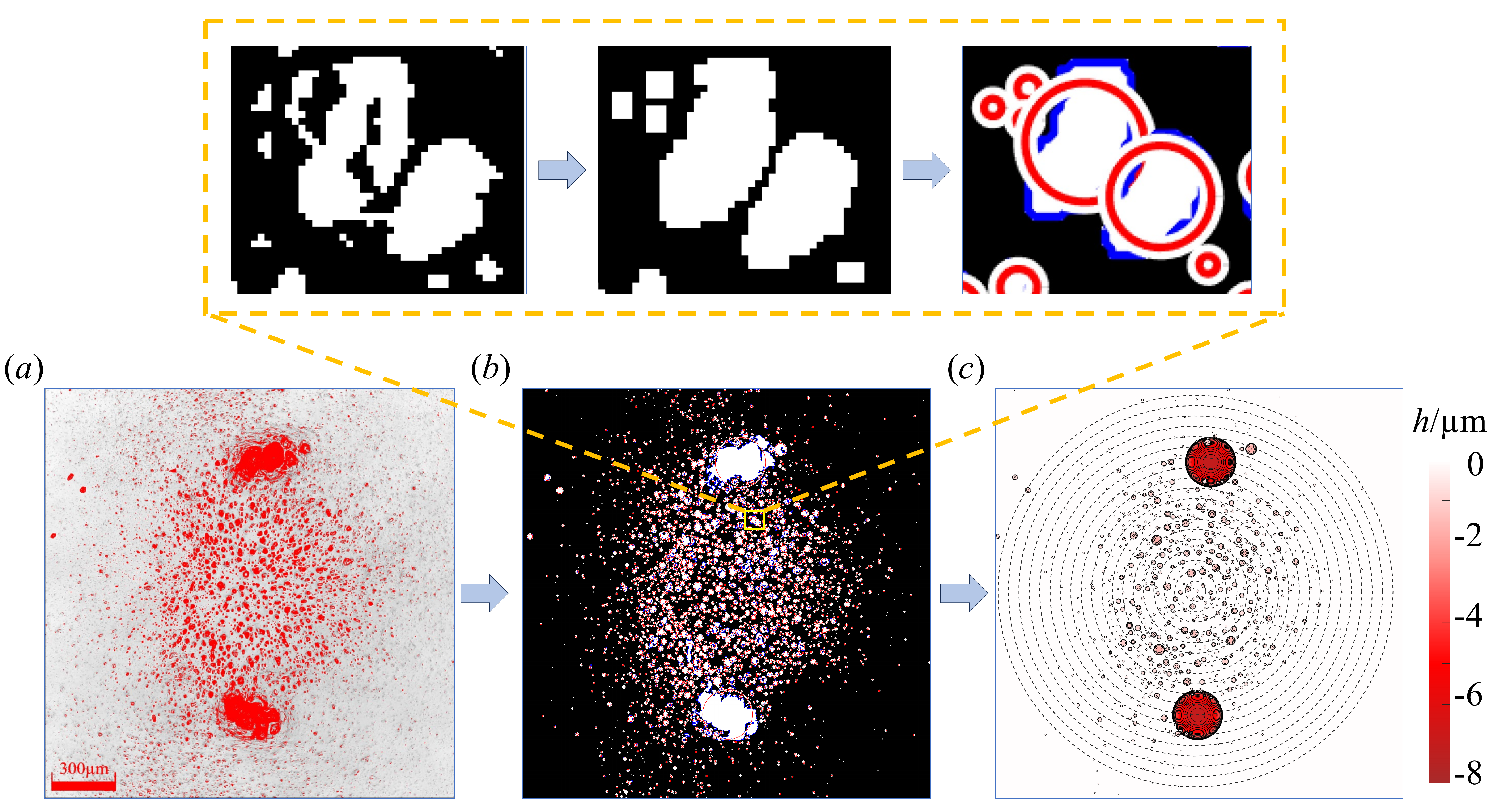}
    \caption{The illustration of the operations to quantify the cavitation erosion. $(a)$ Pixel threshold selection. $(b)$ Binary-based pits image processing with simultaneous edge detection and equivalent circular pits analysis. $(c)$ Contour map of cavitation erosion pit depths.} 
    \label{fig: image-processing}
\end{figure}

\begin{figure}
    \centering
    \includegraphics[width=0.90\linewidth]{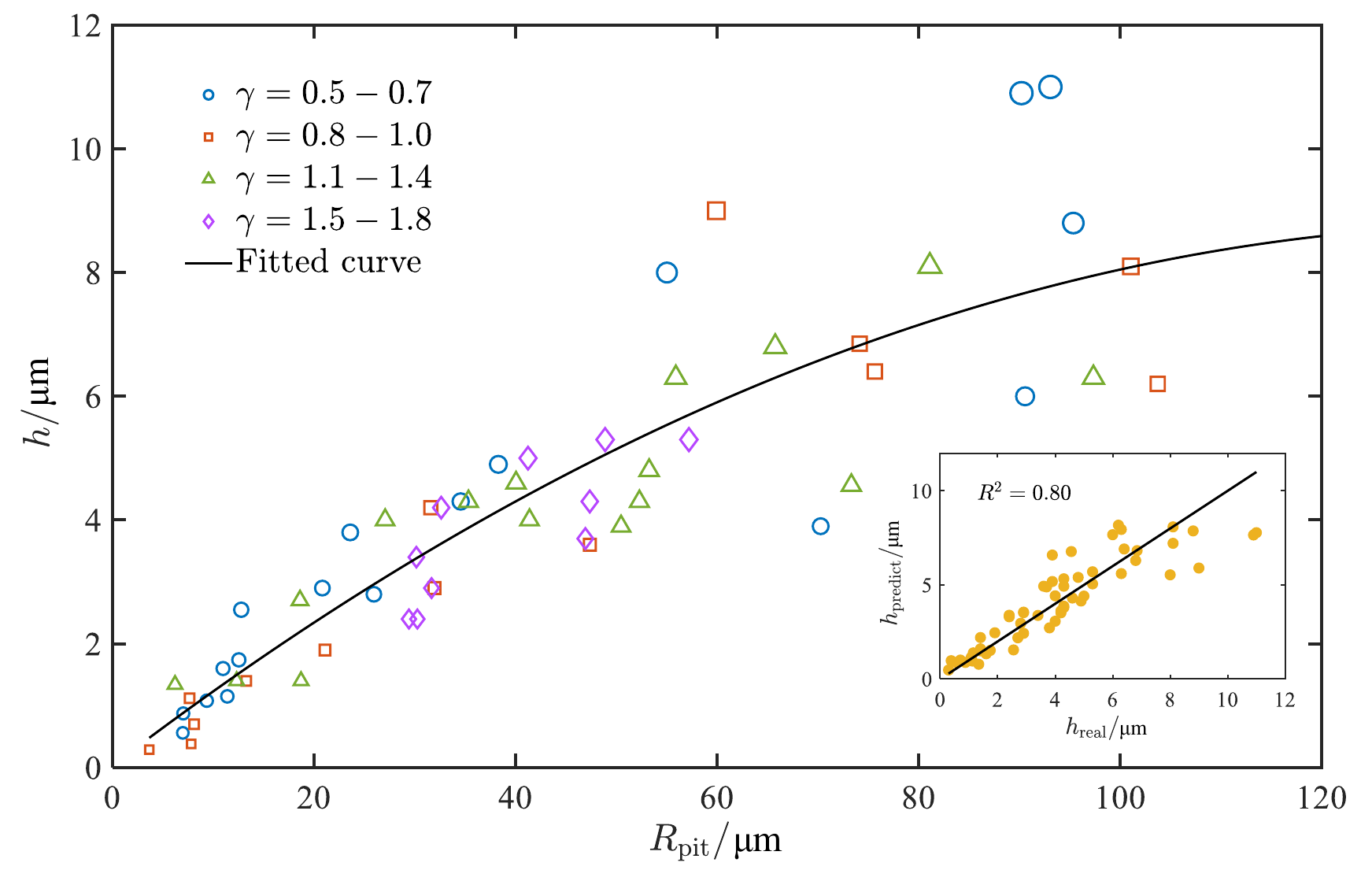}
    \caption{The relationship between the equivalent radius and depth of the pits, regarded as axisymmetric ellipsoid shape, on the aluminium specimens after the 900 times of bubble collapse.} 
    \label{fig:pit_r_d}
\end{figure}

In addition, we also performed a quantitative measurement of the erosion volume for $N = 900$ as discussed in \S\ref{3.1}. 
It should be noted that this study did not directly use the 3D reconstruction function of the microscope to obtain the damaged fields of all specimens for the $N = 900$ cases, as direct 3D measurement is highly time-consuming and has a limited field of view.  
In addition, some specimens exhibited slight overall bending deformation, which could introduce significant errors in direct measurements. 
Taking the typical Bipolar pattern shown in figure \ref{fig:cavitation erosion} as an example, figure \ref{fig: image-processing} illustrates how to obtain the quantitative distribution of cavitation erosion through image-based operations. The detailed method can be found in \ref{appendix A}. 
The specimens were divided into four groups based on different stand-off distances, and different-sized pits were randomly selected as sampling data from each group. The volumes of the pits, $V_{\rm pit}$, were measured using three-dimensional microscope. 
According to \cite{dular2013observations}, erosion pits formed on the surfaces of soft metal materials like pure aluminium after plastic deformation are usually approximately axisymmetric ellipsoids. 
Therefore, the equivalent pit depth can be calculated as $h=3V_{\rm pit}/(4\pi R_{\rm pit}^2)$. 
Figure \ref{fig:pit_r_d} shows the fitted quadratic relationship between the equivalent radius and depth of the axisymmetric ellipsoid pit based on the sampled data: $h = -0.0004R_{\rm pit}^{2} + 0.1244R_{\rm pit} + 0.031$ (in micrometers). 
Regression analysis proves the good predictive ability of this quadratic curve. Finally, a contour map of the cavitation erosion pit depths is provided in figure \ref{fig: image-processing}$(c)$ using the fitted curve. This method offers high detection resolution for small-scale pits. 

Based on the quantitative analysis method, figure \ref{fig:Volume}$(b)$ illustrates the overall variation in erosion damage across moderate stand-off distances, specifically within the range $0.58 \le \gamma \le 2.20$, and compares with the cases $N = 1$. To maintain consistency in the methodology throughout the study, the value of $\gamma$ is calculated using direct measurement of $R_{\max}$. Compared to the $N = 1$ case, the variation trend of cavitation volume with \(\gamma\) for $N = 900$ is generally consistent, although the peak near \(\gamma = 1\) is less pronounced. Moreover, since each specimen underwent hundreds of bubble cycles under the same conditions, random errors were reduced, resulting in significantly lower overall dispersion. 
At $\gamma = 2.2$, the extent of cavitation erosion is already considerably small, and it is expected to further diminish monotonically with further increasing distance. 
Additionally, the fitted curves of the average cavitation erosion volume for both cases exhibit similar shapes. In conjunction with the discussion in \ref{appendix B} regarding the impact of bubble repetition count on the cavitation erosion distribution, it is observed that erosion caused by a single bubble collapse is random. However, as the number of bubbles increases, a statistically consistent erosion pattern begins to emerge, indicating that the bubble evolution process and collapse mode remain consistent across occurrences.

\subsection{Characteristics of bubble collapse}\label{3.3}%实际上现在是3.3了

\begin{figure}
    \centering
    \includegraphics[width=0.85\linewidth]{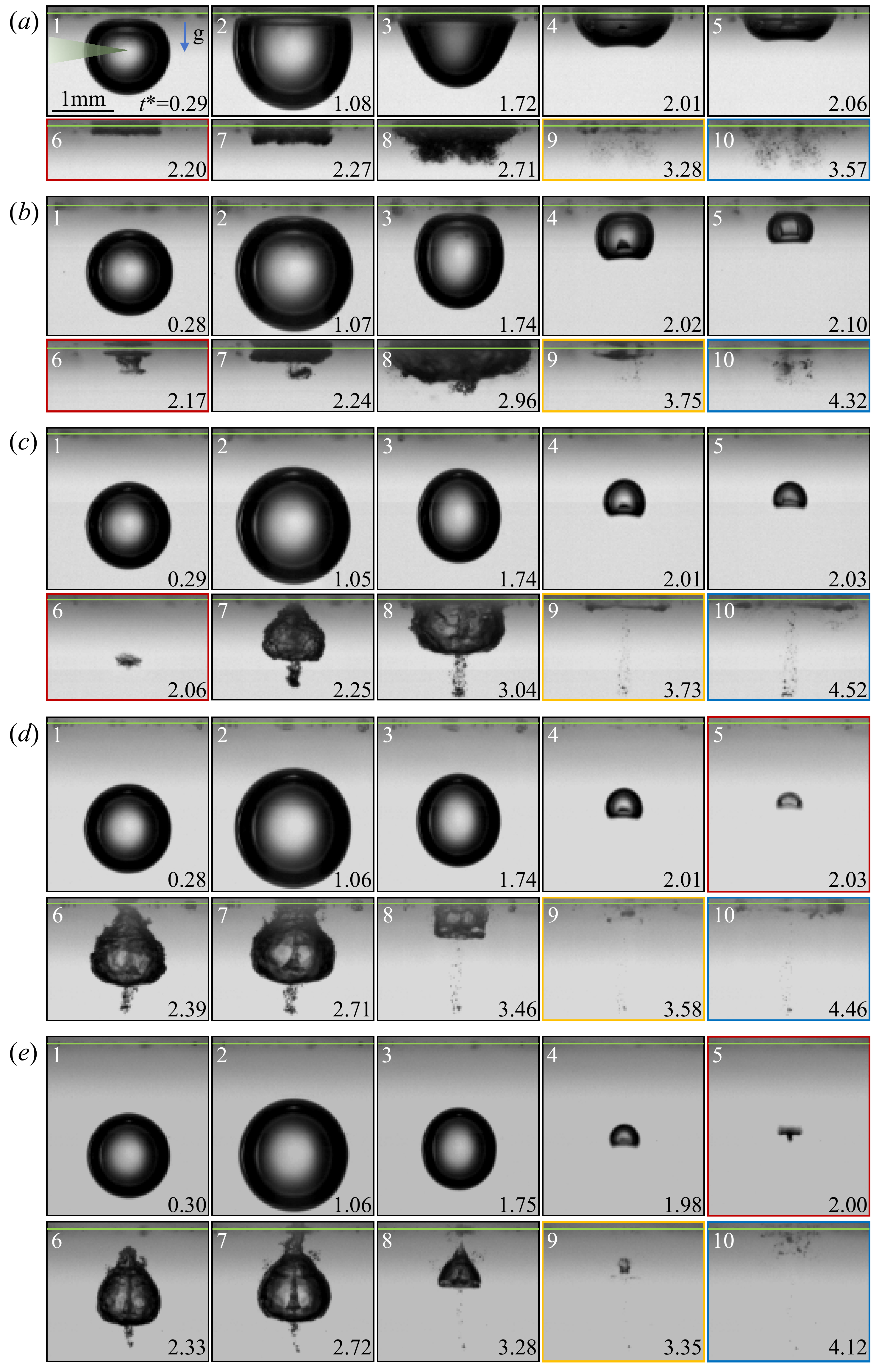}
    \caption{The side-view snapshots of laser-induced bubble at different stand-off distances: (\textit{a}) $\gamma = 0.54$, $R_{\max}$ = 0.96 mm; (\textit{b}) $\gamma = 1.05$, $R_{\max}$ = 0.96 mm; (\textit{c}) $\gamma = 1.50$, $R_{\max}$ = 0.93 mm; (\textit{d}) $\gamma = 1.78$, $R_{\max}$ = 0.95 mm; (\textit{e}) $\gamma = 1.95$, $R_{\max}$ = 0.90 mm. 
    The red, yellow, and blue boxes mark the moments when bubbles collapse to the minimum volume for the first, second, and third time, respectively. The green line above each frame represents the position of the wall. The blue arrow denotes the direction of gravity, and the green light path indicates the direction of laser incidence.
    }
    \label{fig:bubble near wall exp}
\end{figure}

\begin{figure}
    \centering
    \includegraphics[width=0.72\linewidth]{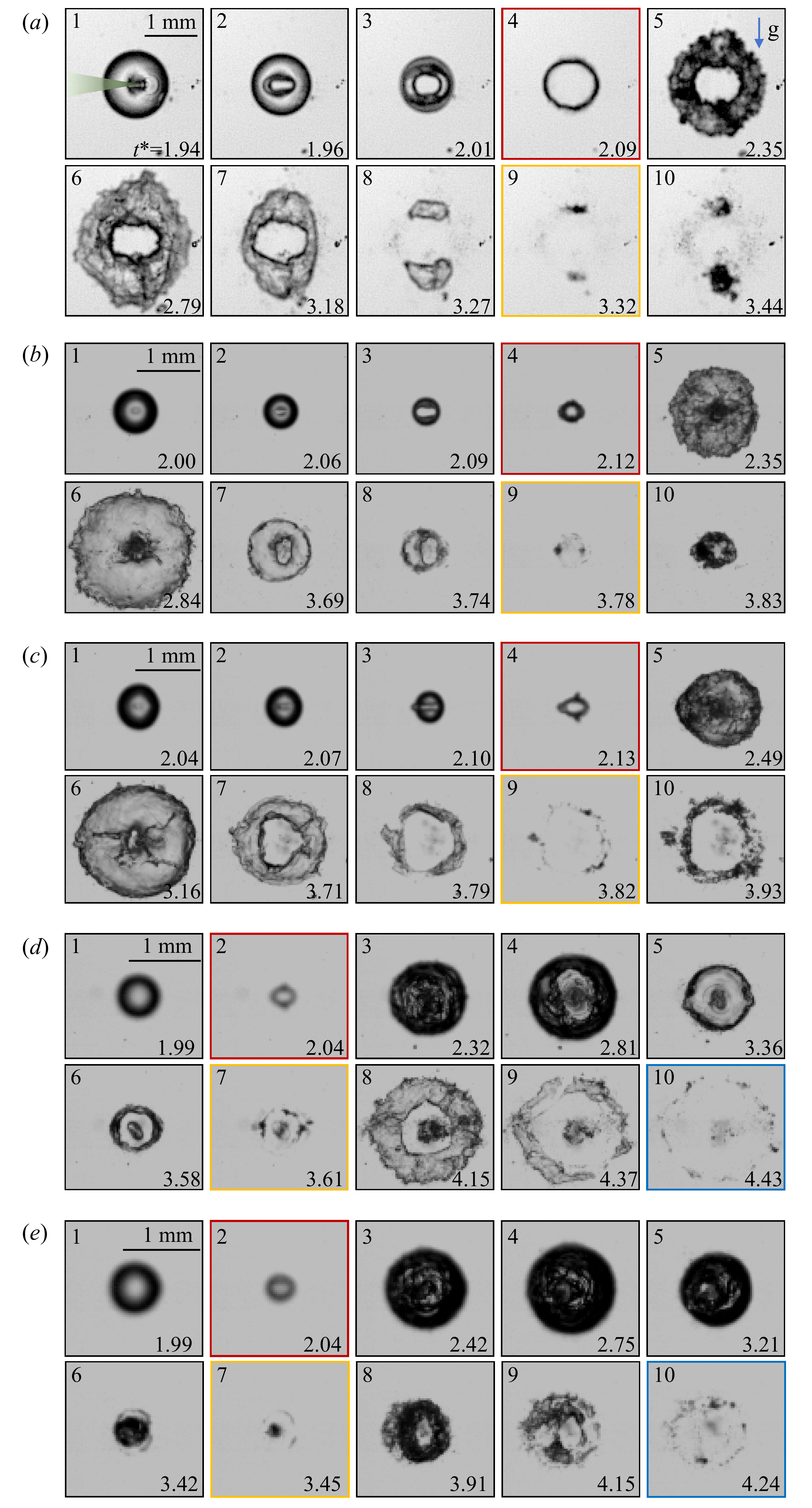}
    \caption{The plan-view snapshots of typical regimes of bubble pulsation near a rigid wall: (\textit{a}) Bipolar, $\gamma = 0.59$, $R_{\max}$ = 0.95 mm. (\textit{b}) Mini-Annular, $\gamma = 1.06$, $R_{\max}$ = 0.79 mm. (\textit{c}) Annular, $\gamma = 1.52$, $R_{\max}$ = 0.83 mm. (\textit{d})  Solar-Halo, $\gamma = 1.77$, $R_{\max}$ = 0.84 mm. (\textit{e}) Central, $\gamma = 1.98$, $R_{\max}$ = 0.85 mm. The red, yellow, and blue boxes mark the moments when bubbles collapse to the minimum volume for the first, second, and third time, respectively. The blue arrow denotes the direction of gravity, and the green light path indicates the direction of laser incidence.
    }
    \label{fig:side view}
\end{figure}

The pulsation characteristics of near-wall cavitation bubbles were investigated, focusing on their collapse modes. 
Figure \ref{fig:bubble near wall exp} presents snapshots of bubble evolution at different instants for bubbles with nearly identical $R_{\max}$ but varying stand-off distances from the wall.
As shown in figure \ref{fig:bubble near wall exp}($a$), the pulsation behavior of bubbles nucleating close to the wall resembles the process schematized in figure \ref{fig:Cavitation erosion diagram}($b$) for bubbles at $\gamma<1$.
However, a notable observation not indicated in the schematic diagram is that the bubble fragmented into a cluster of microbubbles in the later stages of the second collapse. 
For bubbles with a stand-off distance approximately equal to their maximum radius, as depicted in figure \ref{fig:bubble near wall exp}($b$), the bubble reached its minimum volume at the end of the first collapse, generally coinciding with the moment when the jet impacted the wall. 
It is worth noting that, in comparison to the case at $\gamma=0.54$, the bubble at $\gamma=1.05$ did not make direct contact with the wall during its initial collapse, thereby mitigating the wall's hindrance to the radial contraction of the bubble. 
Consequently, this resulted in a more rapid collapse and a smaller final volume of the bubble.
Additionally, the bubble was pierced by the jet and torn into two distinct parts, as shown in frame 6 \citep{philipp1998cavitation, yang2013dynamic}. 
The interaction between these two parts amplifies instability, resulting in a pronounced asymmetry in the bubble pulsation during the second cycle, as shown in figure \ref{fig:side view}($b$).
At $\gamma=1.50$, as illustrated in figure \ref{fig:bubble near wall exp}($c$), the bubble pulsation at this greater stand-off distance corresponds to the dynamics shown in figure \ref{fig:Cavitation erosion diagram}($a$).
Whereas, a unique phenomenon of ``counter jet'' was observed during the second cycle (frames 7–8), where a slender jet ejected the central microbubbles away from the wall \citep{lindau2003cinematographic,zhang2021effect}. 
Subsequently, the toroidal bubble cluster near the wall collapsed, and even after the second and third collapses, it maintained a relatively regular annular shape, with an increasing radius (figiure \ref{fig:side view}$c$).
At $\gamma=1.78$ (figure \ref{fig:bubble near wall exp}$d$), the bubble’s pulsation during the first cycle resembles the dynamics described in the third case. 
However, the bubble exhibited faster contraction and jet formation, reaching a smaller minimum volume at the end of the first cycle. 
During the second collapse, the bubble surface evolved into a "cylindrical shape" and eventually split into two parts in the middle (frame 8). 
The upper portion then contracted towards the center of the wall, experiencing central collapse.
In the subsequent third cycle, the residual microbubbles in the region away from the wall expanded on the wall, forming a larger ring-shaped microbubble cluster and undergoing a third near-wall collapse. 
Further increasing $\gamma = 1.95$ (figure \ref{fig:bubble near wall exp}$e$), the upper portion of the ruptured bubble in the second cycle did not yet contact the wall (frame 9). As the bubble continued to migrate, the microbubble cluster eventually reached the wall in the third cycle (frame 10), but the interaction between the bubbles and the wall weakened.

We further provide the plan view of bubble pulsation under conditions similar to those in figure \ref{fig:bubble near wall exp}, though not captured simultaneously. From figure \ref{fig:side view}, we can clearly distinguish the axial asymmetric collapse process of these cavitation bubbles and categorize the bubble collapse modes. We have identified six distinct collapse modes: Bipolar, Monopolar, Annular, Mini-Annular, Solar-Halo, and Central. 
Specifically, frame 1 of figure \ref{fig:side view}($a$) shows that the jet tip was not axisymmetric when it contacted the wall, but rather flat, with its long axis parallel to the laser beam. 
This indicates that flow instability developed rapidly under small distance parameters, yielding significant asymmetry in the initial stage of bubble collapse in the first cycle.
Subsequently, the jet head becomes teardrop-shaped due to obstruction by the wall (frames 2-3).
Although the toroidal bubble at its minimum volume is regular (frame 4), the surface of the rebounded bubble is much rougher compared to the smooth surface in the first cycle (frames 5-6). As discussed in \cite{plesset1954stability}, this phenomenon was attributed to Rayleigh-Taylor instability.
Moreover, the toroidal bubble exhibited a faster expansion rate in the direction perpendicular to the laser beam, then formed a ``lip-shaped'' bubble during contraction (frame 7). 
Eventually, the ``lip-shaped'' bubble was pinched off, and then the separated sections underwent collapse along the circumferential direction during the second cycle.
Both sections violently contracted to the middle points on the upper and lower ends (frame 9), thus referred to as the Bipolar mode of bubble collapse.
It is obvious that the bipolar bubble collapse is strongly related to the Bipolar pattern of erosion distribution shown in figure \ref{fig:cavitation erosion}.
When the collapse intensity at one end of the ring is much higher than the opposite side, or starts at a point and propagates circumferentially, this asymmetric process is known as the Monopolar collapse mode (see figure \ref{fig: two-side}$b$). This mode frequently occurs and is associated with the Monopolar erosion pattern within the range of $\gamma$ from 0.5 to 0.9. 
If the laser-induced bubble has good initial sphericity, a ‘lip-shaped’ bubble would not form, and the bubble collapse in the second cycle would remain essentially axisymmetric, namely the Annular collapse.
The influence of the initial bubble shape on collapse modes will be detailed later in \S\ref{4.1}.
Figure \ref{fig:bubble near wall exp}($b$) and \ref{fig:side view}($b$) both show the Mini-Annular bubble collapse mode at $\gamma$ approximately equal to 1.0. 
This mode is characterized by the considerably small toroidal bubble radius ($R_s$) formed during the secondary collapse, typically within the range of $R_s < 0.5 R_{\max}$. 
Interestingly, within the range of $\gamma$ dominated by the Mini-Annular mode, the cavitation erosion pattern is mostly Monopolar, and the region where the most severe cavitation occurred is even beyond $R_s$, similar to the findings reported by \cite{wen2023investigation}. The relevant mechanism will be discussed in \S\ref{3.4}.
When $\gamma$ is around 1.50, as shown in figure \ref{fig:side view}($c$), the toroidal bubble splits into a cluster of microbubbles during the second contraction cycle, they are generally uniformly distributed along the circumferential direction, corresponding directly to the annular erosion pattern.
By comparing figure \ref{fig:side view}($d$) and ($e$), similarities can be observed between the Solar-Halo mode and the Central mode during the first and second cycles.  
However, for the Solar-Halo mode, damage in the ``solar'' region was caused by the bubble collapse in the second cycle, while the collapse of the third cycle led to damage in the ``halo'' region.
On the other hand, the Central mode exclusively caused damage in the center as a result of the collapse of the second cycle, and the impact of the third collapse on the wall surface can be disregarded.

In general, it can be concluded that there is a strong correlation between bubble collapse modes and the associated cavitation erosion patterns, except that Monopolar distribution of damage can also be attributed to bubble collapse of the Mini-Annular mode, in addition to bubble collapse of the Monopolar mode.

\subsection{Cavitation erosion mechanisms}\label{3.4}

\begin{figure}
    \centering
    \includegraphics[width=1\linewidth]{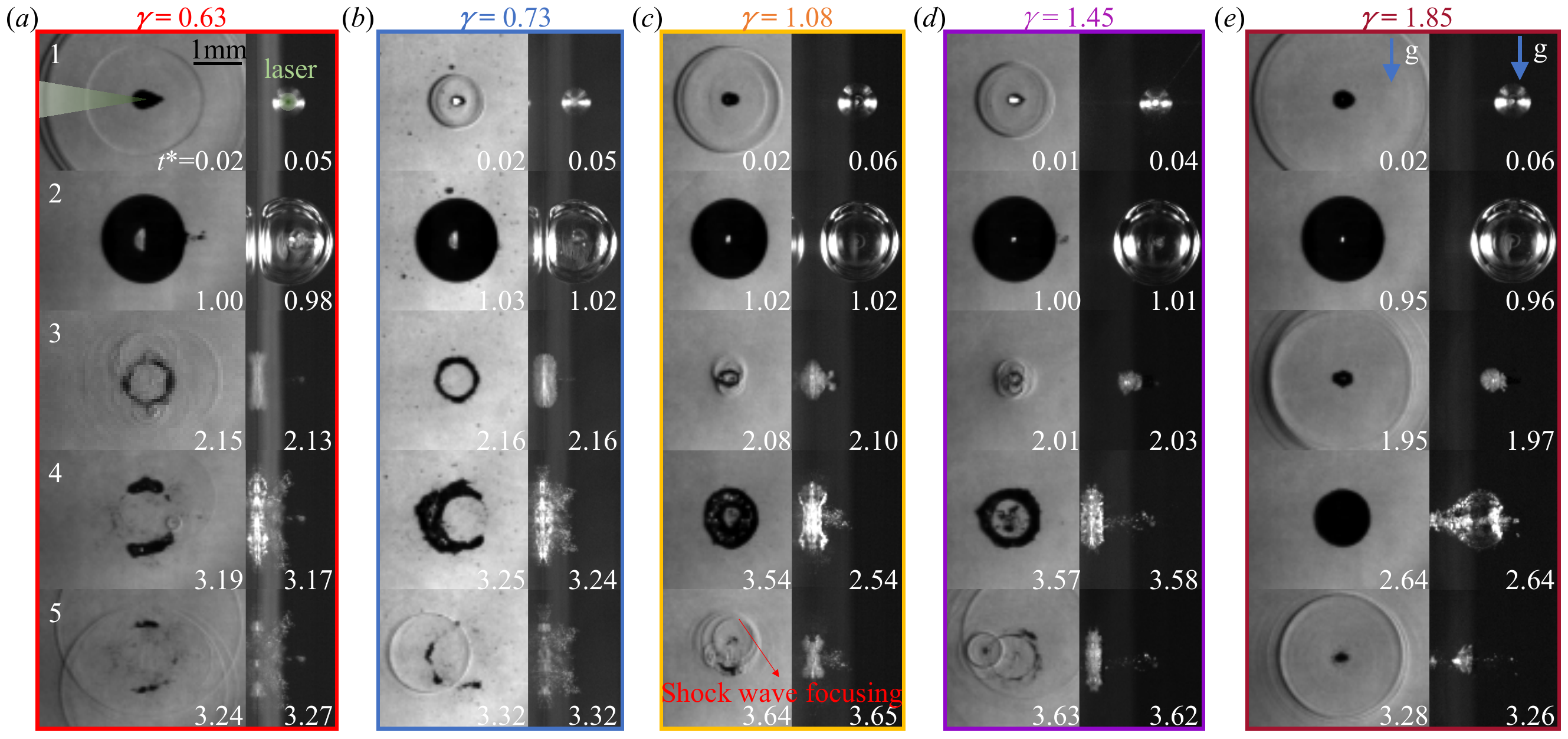}
    \caption{Comparison of shockwave front propagation caused by bubble collapse at different $\gamma$. Each set consists of the plan-view schlieren images on the left column and side-view snapshots on the right column. The bubble collapse modes and characteristic parameters are: $(a)$ Bipolar, $\gamma = 0.63$, $R_{\max} = 1.01$ mm; $(b)$ Monopolar, $\gamma = 0.73$, $R_{\max} = 0.97$ mm; $(c)$ Mini-Annular, $\gamma = 1.08$, $R_{\max} = 0.92$ mm; $(d)$ Annular, $\gamma = 1.45$, $R_{\max} = 0.97$ mm; $(e)$ Central, $\gamma = 1.85$, $R_{\max} = 0.98$ mm. The blue arrow denotes the direction of gravity, and the green light path indicates the direction of laser incidence. The green dot indicates the laser path is perpendicular to the paper, pointing outward.}
    \label{fig: two-side}
\end{figure}

\begin{figure}
    \centering
    \includegraphics[width=1\linewidth]{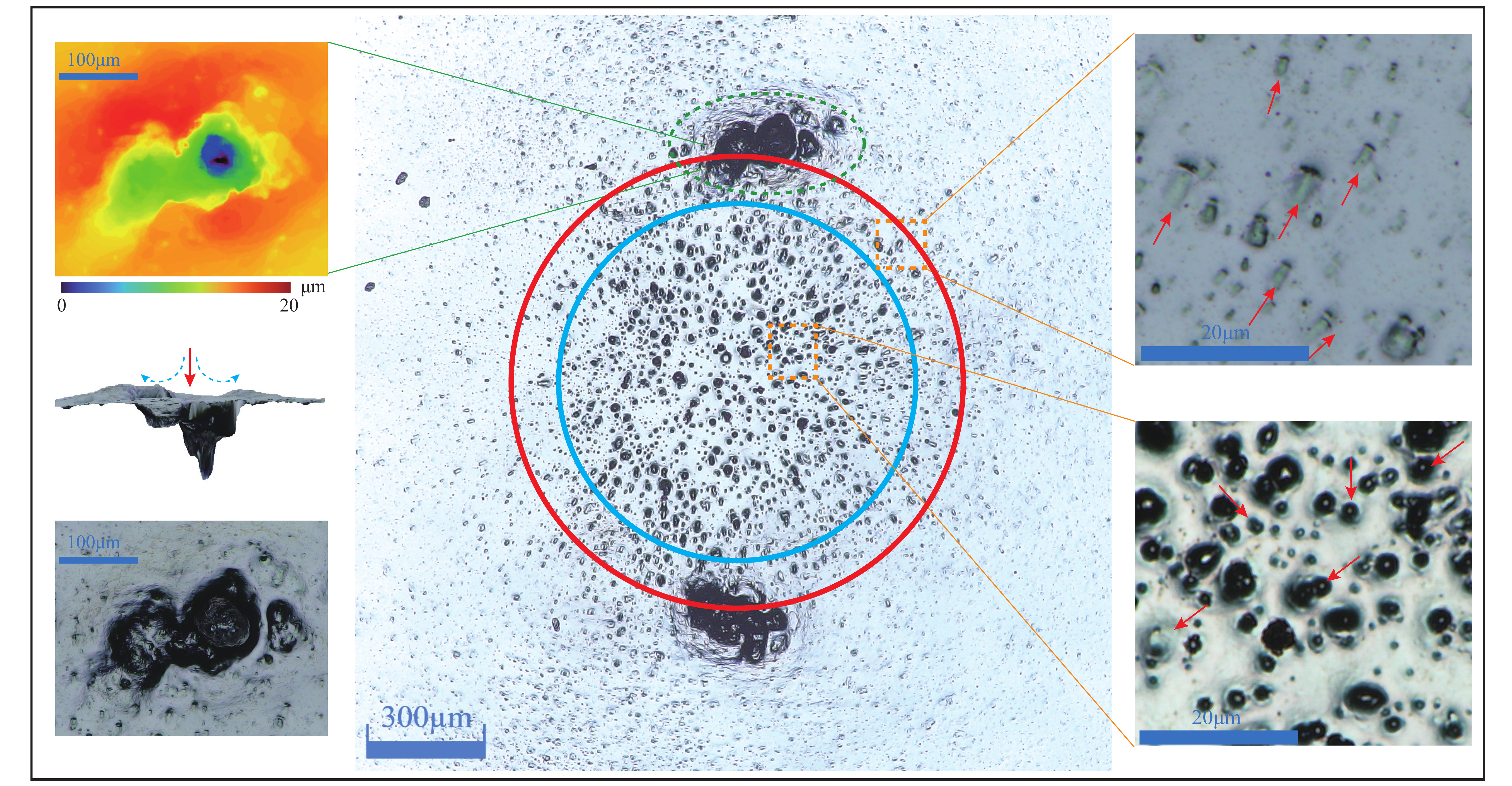}
    \caption{ The close-up views of the Bipolar pattern of erosion damage. The blue and red circles represent the position of the toroidal bubble at the end of the first and second collapses, respectively.
    The red arrows indicate the load direction inferred from the plastic deformation of the material surrounding the pit. 
    }
    \label{fig: erosion-mechanism}
\end{figure}

\begin{figure}
    \centering
    \includegraphics[width=1\linewidth]{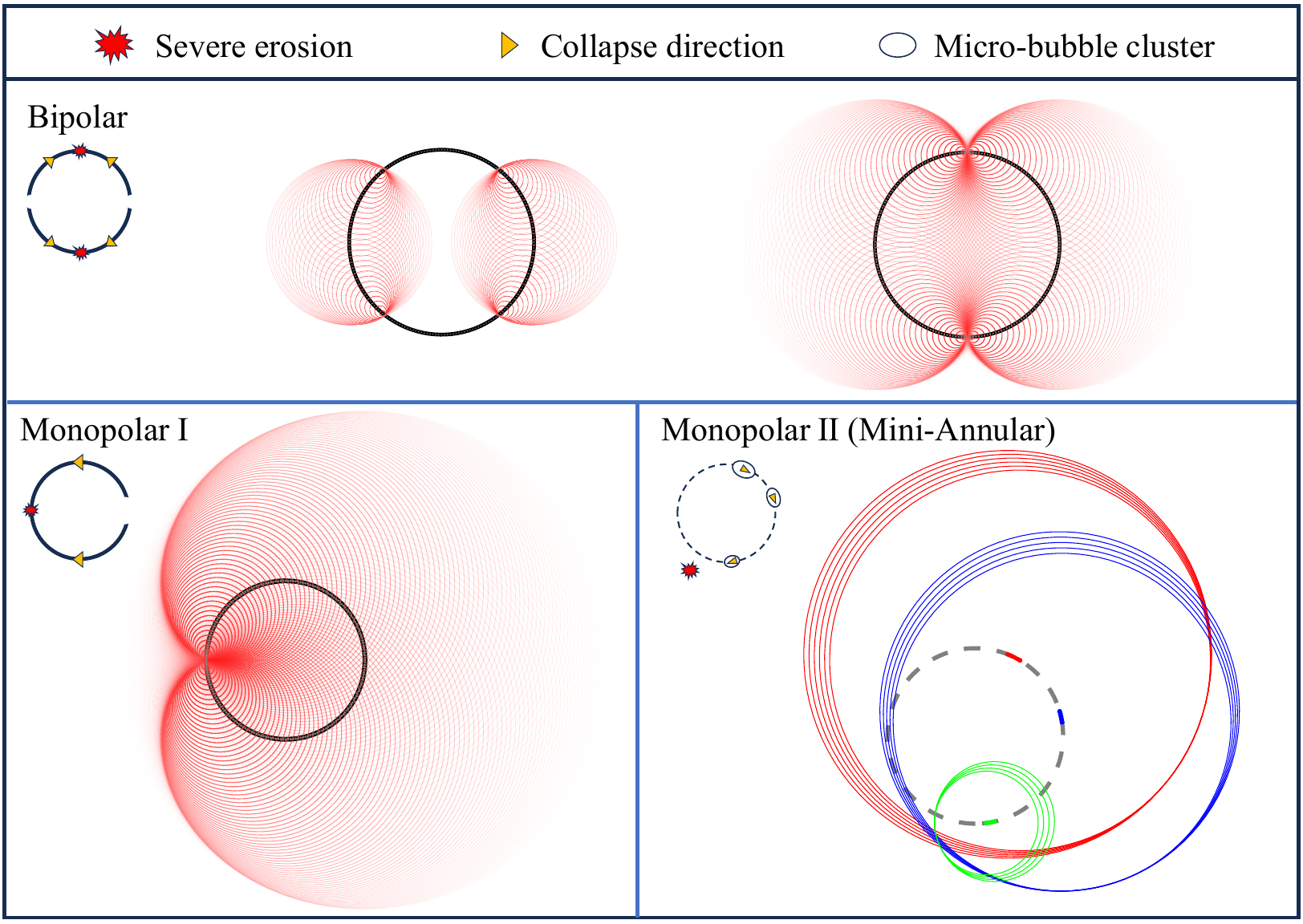}
    \caption{Schematic diagram of the shockwave focusing mechanisms on inducing severe erosion damage in the Bipolar and Monopolar patterns.
    The black circular rings indicate the toroidal bubble at the second cycle, while the grey dashed ring of the Monopolar \uppercase\expandafter{\romannumeral2} (Mini-Annular) indicates the separated parts of the annular microbubble cluster.}
    \label{fig: shock-wave-diagram}
\end{figure}

We demonstrate the link between bubble collapse modes and cavitation erosion patterns, thereby revealing the erosion mechanism at moderate stand-off distances. Initially, we analyze the shock waves produced by four distinct bubble collapse modes through the Schlieren images presented in figure \ref{fig: two-side}. The series of five images captures the initial condition of each bubble (frame 1), key moments in the first cycle (frames 2 \& 3) and the second cycle (frames 4 \& 5).

Figure \ref{fig: two-side}$(a)$ demonstrates that in the Bipolar mode, the axially symmetric collapse of a toroidal bubble during the initial cycle resulted in asynchronous shock waves, forming irregular wavefronts due to complex wave interfence.
On the other hand, during the second cycle, the asymmetric lip-shaped bubble collapsed along the circumferential direction, generating pronounced shock waves with sharp wavefronts originating from two polar regions. 
As depicted in figure \ref{fig: erosion-mechanism}, these two areas precisely correspond to the regions experiencing the most intense cavitation erosion, situated at the upper and lower portions of the red circle. 
Inspired by \cite{reuter2022cavitation}, the schematic in figure \ref{fig: shock-wave-diagram} clarifies the shockwave focusing mechanism accountable for this effect. The evenly spaced points along the toroidal bubble represent the emission sources of shock waves induced by bubble collapse. It is assumed that the spherical shock waves propagate at the speed of sound ($c$ = 1500 m/s) and that the bubble’s circumferential collapse velocity is also $c$. 
Initially, the collapse of a bubble shaped like a lip, propagating circumferentially toward the middle at each end. The subsequent shockwave fronts met and superimposed along the direction of the bubble collapse,  forming a reinforced wavefront with increased pressure.
Furthermore, when the composite wavefronts propagating in opposite directions meet at the middle points, their head-on collision generates a more intense detonation wave focusing, resulting in significantly higher localized pressures at the middle points. 
As illustrated in figure \ref{fig: erosion-mechanism}, the stress imposed on the wall due to the focused shock wave far exceeded the yield strength of pure aluminium, resulting in extensive plastic deformation. 
This deformation manifests as deep pits with diameters greater than 100 $\upmu$m, along with adjacent wrinkled protrusions. 
These characteristics indicate that the deformation caused by the impact at high strain rates induced the plastic flow of aluminium, resembling a liquid-like state. 
Two subfigures located on the right side of figure \ref{fig: erosion-mechanism} illustrate that the deformation of small-scale pits outside the blue circle predominantly pointed radially outward. 
This suggests that the collapse of the toroidal bubble during the first cycle is the primary cause of these pits. 
However, the deforming direction of small-scale pits inside the blue circle seems more variable. 
We posit that this variability stems from the complex wave interference inside the toroidal bubble. The animation to further illustrate the cavitation erosion mechanism for the Bipolar pattern is provided in the supplementary movie \textcolor{red}{1}. 

The Monopolar erosion patterns, as shown in figure \ref{fig:side view}, are marked by intense cavitation erosion localized to one side of the toroidal bubble.
This pattern occurs frequently over a wide range of $\gamma$ values, approximately $0.7 \sim 1.3$. 
Unlike the Bipolar pattern, which exhibits mirror symmetry, the Monopolar pattern breaks this symmetry and may be driven by distinct mechanisms.
The first mechanism is similar to the shock wave focusing mechanism observed in the Bipolar collapse. 
As shown in figure \ref{fig: two-side}$(b)$, during the second cycle, the toroidal bubble undergoes a circumferentially asymmetric collapse, with a higher curvature on the right side of the ring compared to the left (frame 4). As a result,  the right side of the toroidal bubble was pinched off first, leading to a crescent-shaped bubble, i.e. Monopolar collapse mode. 
As illustrated in the schematic (Monopolar \uppercase\expandafter{\romannumeral1}) in figure \ref{fig: shock-wave-diagram}, the shock wavefronts superimposed along the collapse direction of the toroidal bubble and ultimately converged at the left side of the toroidal bubble, inducing the residual bubble to release a strong shock wave (see figure \ref{fig: two-side}$b$, frame 5). 
The second mechanism typically appears in the Mini-Annular collapse mode. 
During the contraction phase of the second cycle, the toroidal bubble fragments into multiple discrete small-scale bubbles. 
These small bubbles then collapsed and emitted shock waves (figure \ref{fig: two-side}$c$).
The wavefronts underwent oblique superposition and concentrated in specific regions outside the toroidal bubble, as indicated in figure \ref{fig: shock-wave-diagram} (Monopolar \uppercase\expandafter{\romannumeral2}). 
This kind of shockwave focusing leads to a lower energy density compared to the first mechanism, thus the overall damage volume in the $\gamma$ range influenced by this mechanism is relatively modest, as depicted in figure \ref{fig:Volume}.
At greater distances, the Monopolar pattern can also be observed as shown in figure \ref{fig:cavitation erosion} (frame 4).
Under these conditions, the toroidal bubble fragmented into several larger sections due to instability (figure \ref{fig:side view}$c$ and \ref{fig: two-side}$d$). 
Upon the bubble collapse, the two aforementioned mechanisms (head-on collision of convergent wavefronts and oblique superposition of wavefronts) may act simultaneously to generate strong shock-wave load. As a result, a considerable amount of cavitation erosion volume is apparent in the $\gamma$ range from 1.3 to 1.5.

Overall, the loads that cause severe cavitation erosion at moderate distances are not generated in the first cycle but arise from the circumferentially asymmetric collapse of the toroidal bubble during the second cycle.

\subsection{Radial distribution of erosion depth}\label{3.5}

\begin{figure}
    \centering
    \includegraphics[width=0.94\linewidth]{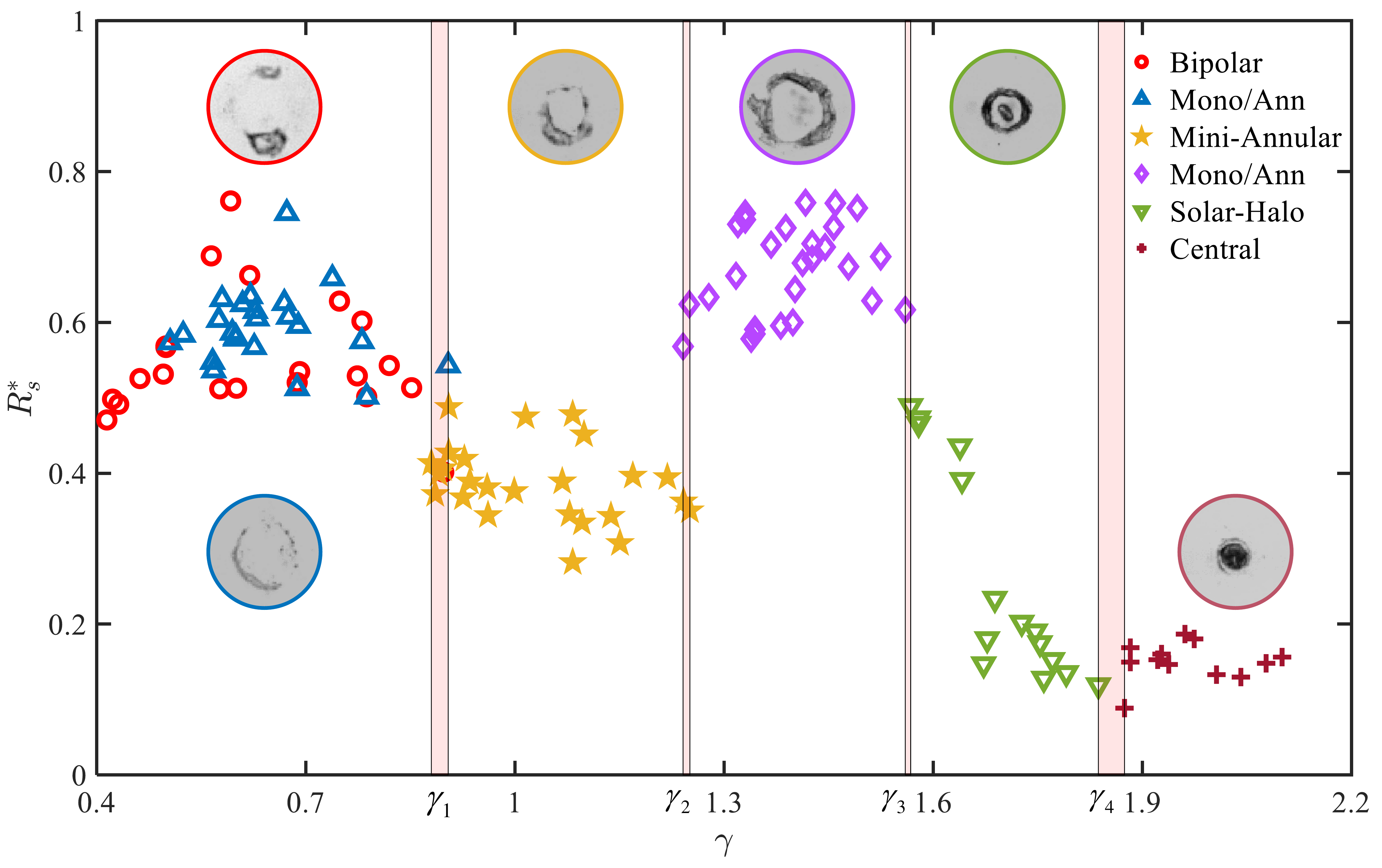}
    \caption{Variations in the normalised radius of the annual bubble in the second cycle, $R^*_{s}=R_s/R_{\max}$, with respect to $\gamma$.
    The collapse modes are marked, with typical examples shown accordingly. Critical stand-off distance parameters for bubble collapse mode transitions: $\gamma_1 \approx 0.89 $, $\gamma_2 \approx 1.25 $, $\gamma_3 \approx 1.56 $, $\gamma_4 \approx 1.85$.}
    \label{fig: Rs}
\end{figure}

In this subsection, we investigate the radial distribution characteristics of cavitation erosion using the same aluminum plate samples discussed in \S\ref{3.1}.
Figure \ref{fig: Rs} presents the normalized outer radius of the toroidal bubble, $R^*_s$, and its corresponding collapse mode during the second collapse phase for $0.4 < \gamma <2.2$, with initial bubble shapes having an aspect ratio, $\eta$, ranging from 1.1 to 1.8. 
Firstly, the Bipolar, Monopolar, and Annular modes coexist for $\gamma$ values between 0.4 and 0.8.
Meanwhile, $R^*_s$ initially increases and then decreases, reaching a peak at approximately 0.6. 
The common feature observed in this range is that the jet impacts the wall before the bubble fully collapses in the first cycle. 
The interaction between the jet and the wall contributes to the formation of a stable vortex ring, resulting in the toroidal bubble maintaining its intact and continuous shape during the first cycle and enlarging its diameter during the second cycle.  
In the region dominated by the Mini-Annular mode of bubble collapse ($0.89<\gamma<1.25$), $R^*_s$ is relatively small.
This is because the jet impact and collapse to minimum volume occur almost simultaneously in the first cycle before the toroidal bubble is fully formed, leading to a weaker vortex intensity. 
As the toroidal bubble is already in close proximity to the wall during the second cycle and does not generate strong jets to expand its inner diameter, $R^*_s$ ends up being small. 
In contrast, during the third range ($1.25<\gamma<1.56$), toroidal bubbles that collapse in the Mono/Ann modes exhibit higher-intensity jets during the second cycle. 
These jets, blocked by the wall, result in a radial outward flow that enlarges the inner diameter of the toroidal bubble and contributes to the appearance of a second peak in $R^*_s$. 
At $\gamma>1.6$, most of the potential energy of the bubble is dissipated during the first two collapses, leading to a monotonic decrease in the radius of the bubble ring with increasing $\gamma$, corresponding to the Solar-Halo mode.  
Finally, in the range of $1.8<\gamma<2.2$, a transition occurs to the Central mode, characterized by the smallest $R^*_s$ values.

\begin{figure}
    \centering
    \includegraphics[width=0.95\linewidth]{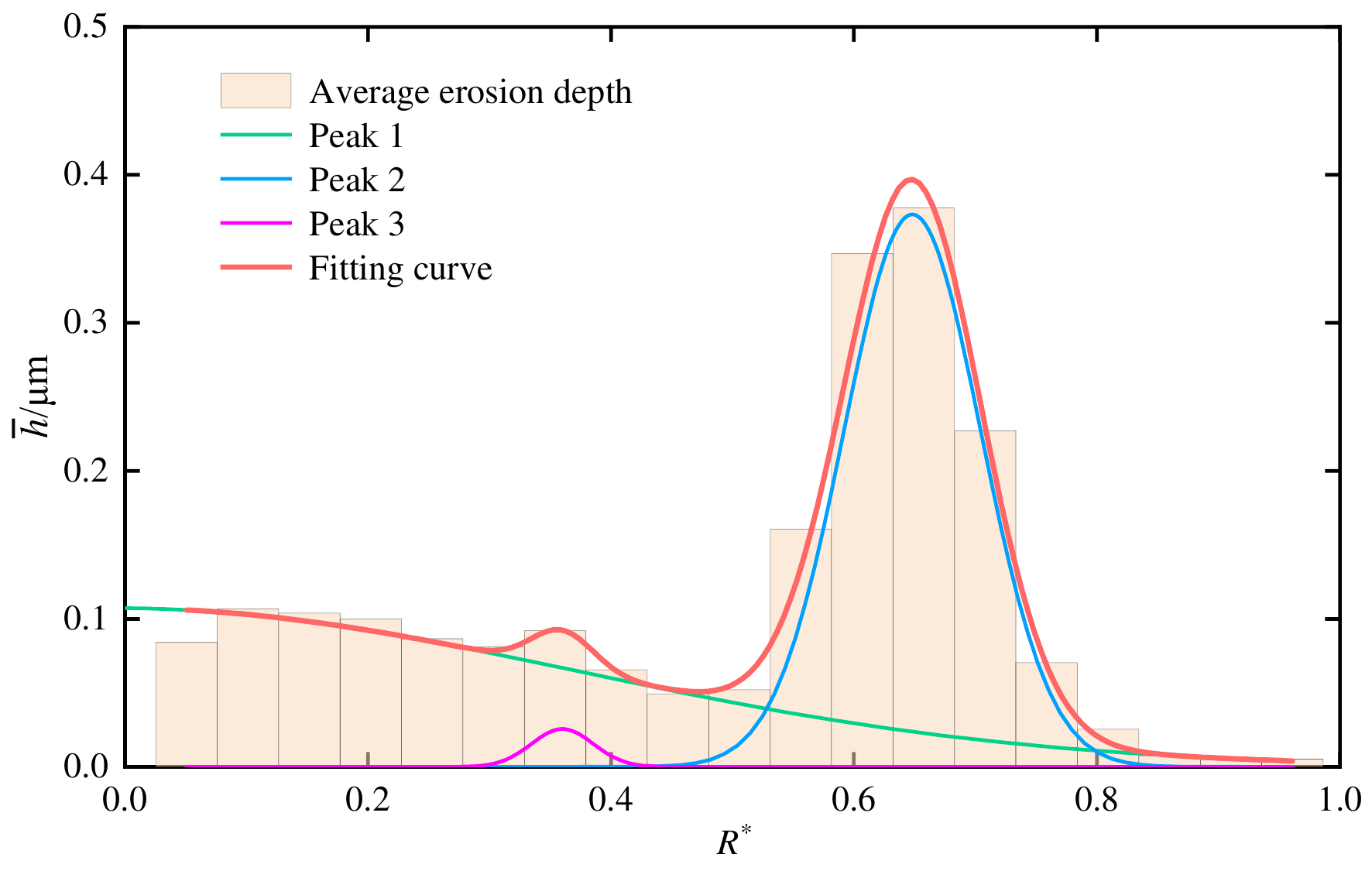}
    \caption{The radial distribution of average cavitation erosion depth at $\gamma = 0.58$. A Gaussian multi-peak fitting curve is provided, with each Gaussian function for the corresponding peak presented. 
    }
    \label{fig: CER}
\end{figure} 

\begin{figure}
    \centering
    \includegraphics[width=1\linewidth]{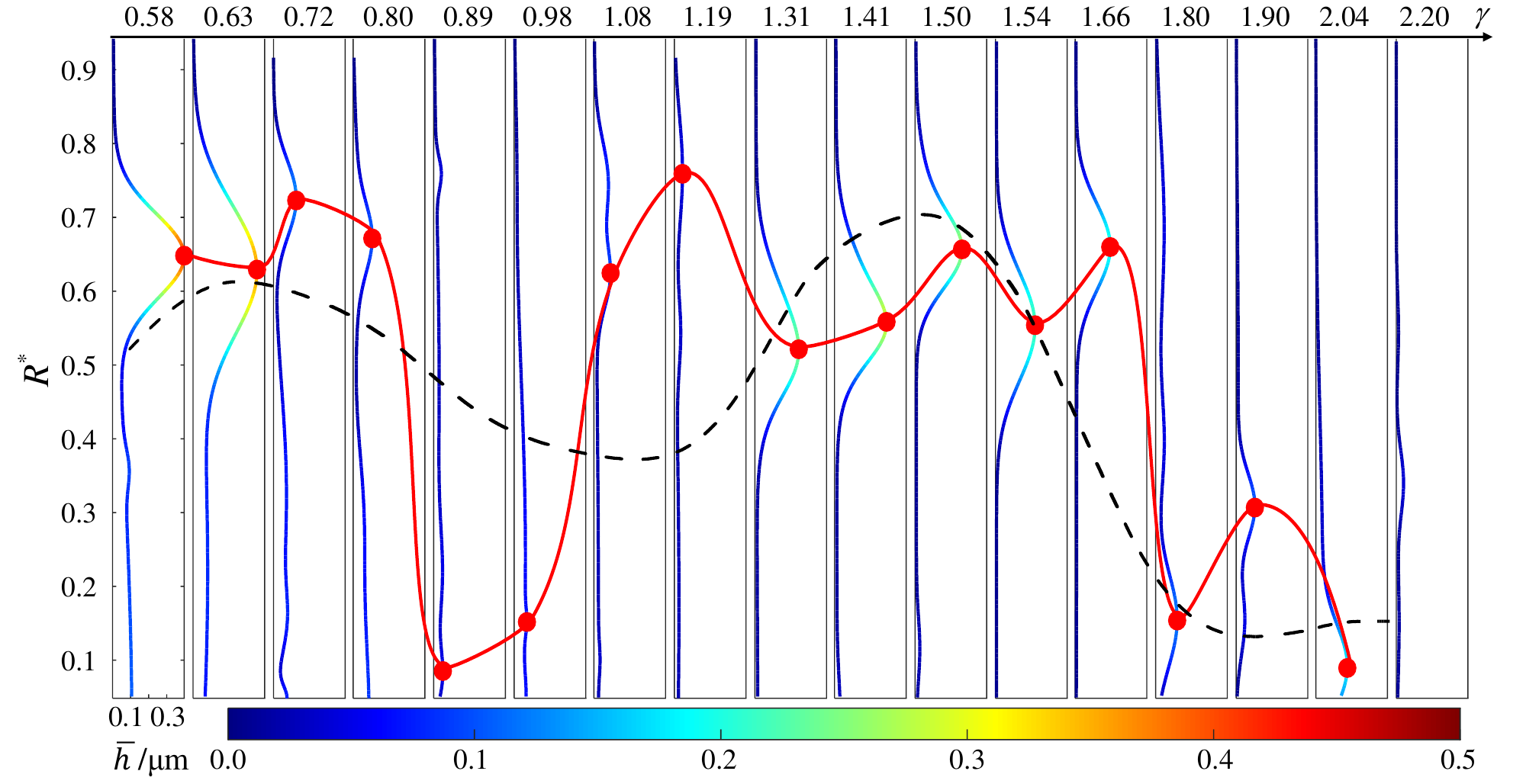}
    \caption{The profiles of radial distribution of cavitation erosion depth from $0.58\le\gamma\le2.20$.
    The dashed line shows the fitted curve of $R^*_{s}$.
    The red circles indicate the radial positions with the highest erosion depth at each $\gamma$.
    }
    \label{fig:CER_Rs}
\end{figure}

Next, we quantitatively analyze the radial distribution characteristics of average erosion depth using the image-based pit reconstruction method proposed in \S\ref{3-2}, as shown in figure \ref{fig: image-processing}. The specific equation used is to obtained average erosion depth:
\begin{equation} \overline{h}_i = \frac{\int^{R_i}_{R_{i-1}}2\pi hRdr}{\pi(R^2_{i}-R^2_{i-1})}, \quad (i=1,2,3, \ldots, n).
\end{equation}
Taking the $\gamma = 0.58$ case as an example, figure \ref{fig: CER} demonstrates the corresponding radial distribution of cavitation erosion using a multi-modal Gaussian function to fit the discrete values of $\overline{h}_i$ $(i=1,2, \ldots, 20)$.
The fitted curve exhibits three peaks, located at $R^*=0$, 0.35, and 0.65, respectively. 
The first two smaller peaks primarily result from the impact load of the jet and toroidal bubble collapse in the first cycle, while the largest peak arises from the shock wave focusing resulting from circumferentially asymmetric collapse in the second cycle. 
Similarly, the fitted curves for the radial distribution of cavitation are plotted for $\gamma =$ 0.58 to 2.20 in figure \ref{fig:CER_Rs}. 
To facilitate comparison, the positions of $R^*_s$ and the maximum erosion depth are marked. 
Overall, for radial positions of $R^*>0.9$, the erosion depth tends to approach zero, indicating that the cavitation erosion effect mainly occurs within the area where the maximum bubble radius projects onto the wall. 
There are almost no noticeable erosion pits on the sample surface at $\gamma = 2.2$, so it is defined as the critical distance for cavitation erosion. 
When $\gamma \le 0.8$, the maximum cavity depth is around $R^*=0.6\sim 0.7$, further substantiating the explanation of the Bipolar cavitation erosion mechanism in \S\ref{3.3}. 
In addition, the position of the maximum cavity when $\gamma = 0.72\sim 0.8$ is slightly larger than the radius of the toroidal bubble. 
This phenomenon can be attributed to the combined effects of two shockwave focusing mechanisms: i) the collision of convergent shockwave front and ii) the oblique superposition of shock waves.
At $\gamma \approx 0.95$, the erosion level reaches a local maximum in the central region, although its absolute value is relatively small. 
This is the result of the shockwave focusing within the small toroidal bubble. 
For $\gamma$ in the range of 1.1 to 1.2, the toroidal bubble collapses in the Mini-Annular mode, forming a Monopolar cavitation erosion pattern due to the second mechanism of shockwave focusing. The maximum erosion location is significantly greater than $R_s$, as described in \S\ref{3.4}.  
Within the range of $\gamma =$ 1.3 to 1.5, the erosion depth is significant and the position generally coincides with $R_{s}$.  
Within the $\gamma$ range, the Monopolar pattern is still primarily caused by the first shockwave focusing mechanism. 
When $\gamma > 1.54$, we found that the most severely damaged region coincides with the position $R_s$, indicating that the Annular collapse mode is the primary cause of the uniformly distributed pits along the ring. 
At $\gamma = 1.66$, the bubble collapse in the Solar-Halo mode causes relatively severe cavitation erosion damage at a position far larger than $R_{s}$, indicating that the third collapse of the bubble (figure \ref{fig:side view}$d$, frame 10) begins to influence the distribution characteristics of the pits. For $\gamma > 1.8$, the bubble remains above the wall during the first two pulsation cycles. Its collapse releases spherical shock waves, significantly dissipating energy in the process (figure \ref{fig: two-side}$e$). As a result, cavitation erosion pits predominantly in the central region of the wall, with the peak average depth being less than 0.1 $\upmu$m.

In conclusion, the radial distribution of cavitation erosion varies significantly at different stand-off distances. 
However, the regions where severe erosion occurs are mostly located near the toroidal bubble during the second cycle. Table \ref{tab:label 2} summarizes the correspondence between bubble collapse modes and erosion patterns, and lists the salient features associated with each pattern.

\begin{table}
\centering
\setlength{\tabcolsep}{2mm}
\renewcommand{\arraystretch}{1.5} % 调整全表的行间距

\begin{tabular}{cccccccc}

 Collapse mode  &Erosion pattern      & Feature                                     \\
\hline % 可以添加一条水平线来明确区分头部和其他行
Bipolar &  Bipolar                            & The most severe erosion at the opposite sides. \\%[3mm]  在这里增加额外的空间
Monopolar; Mini-Annular &  Monopolar                        & The most severe erosion at one side.                                            \\
Annular & Annular                          & Almost annular distribution of pits.                                                 \\
Solar-Halo & Solar-Halo                  & Pits at the center and circumference.                                            \\
Central & Central                           & Concentration of pits only at the central area. \\
\hline
\end{tabular}
\caption{Correspondence between bubble collapse modes and erosion patterns with typical distribution descriptions.}
\label{tab:label 2}
\end{table}

\section{The dynamics of circumferential asymmetrical bubble collapse}\label{4}

\begin{figure}
    \centering
    \includegraphics[width=1\linewidth]{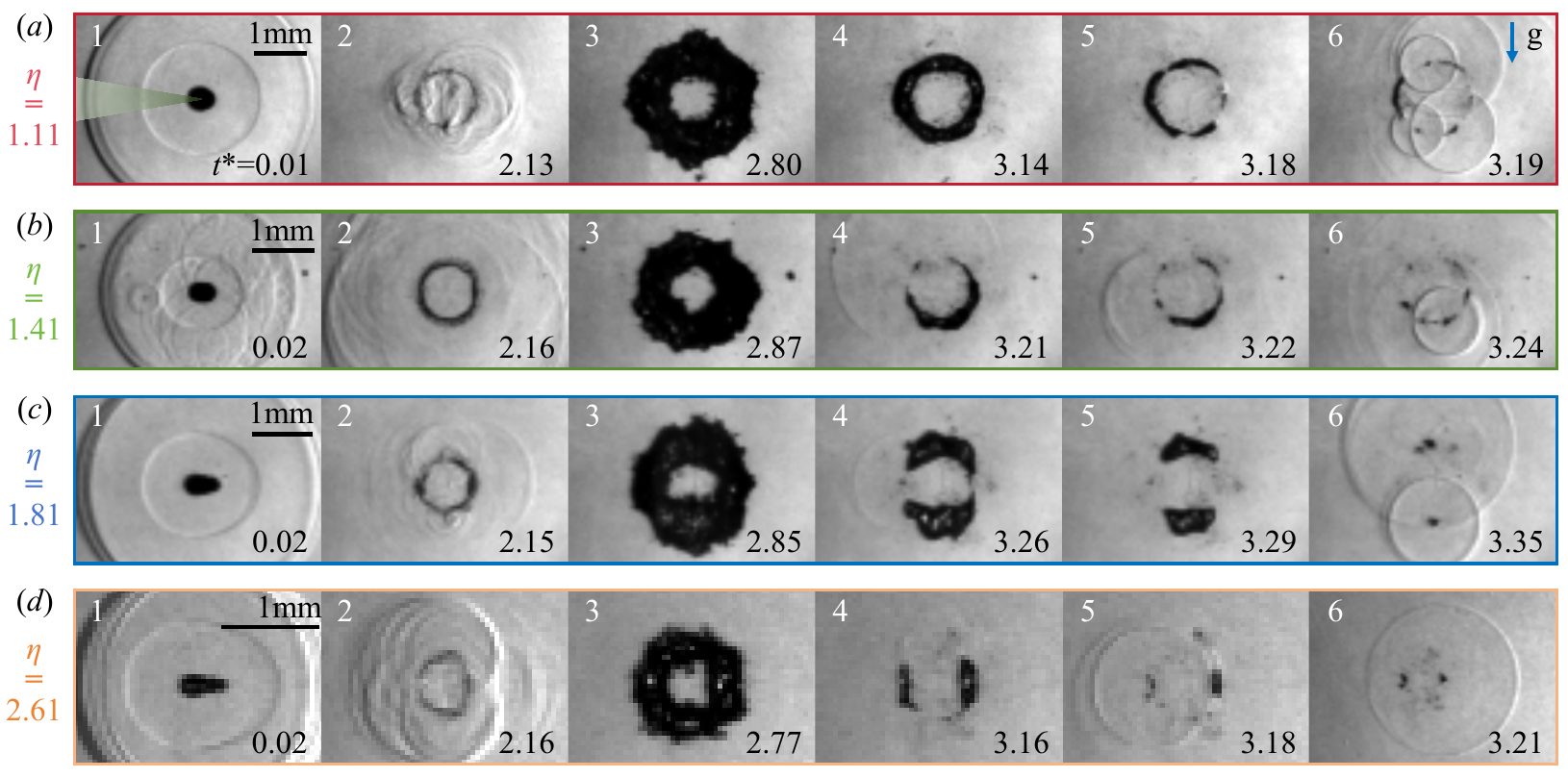}
    \caption{The snapshots of bubbles with different initial aspect ratios $\eta$ within $\gamma=0.60\sim0.70$:
    ($a$) Annular, $\eta=1.11$, $\gamma=0.61$;  ($b$) Monopolar, $\eta=1.41$, $\gamma=0.69$; ($c$) Bipolar-V (the optical path is vertical relative to the line connecting the poles), $\eta=1.81$, $\gamma=0.69$; ($d$) Bipolar-H (the optical path is horizontal relative to the line connecting the poles), $\eta=2.61$, $\gamma=0.61$. 
    The value of $\eta$ is measured using the instantaneous image immediately after plasma flash (frame 1). The second cycle for each one starts at frame 2 and ends at frame 6. The blue arrow denotes the direction of gravity, and the green light path indicates the direction of laser incidence.} 
    \label{fig:eta-shock-wave}
\end{figure}

\begin{figure}
    \centering
    \includegraphics[width=1\linewidth]{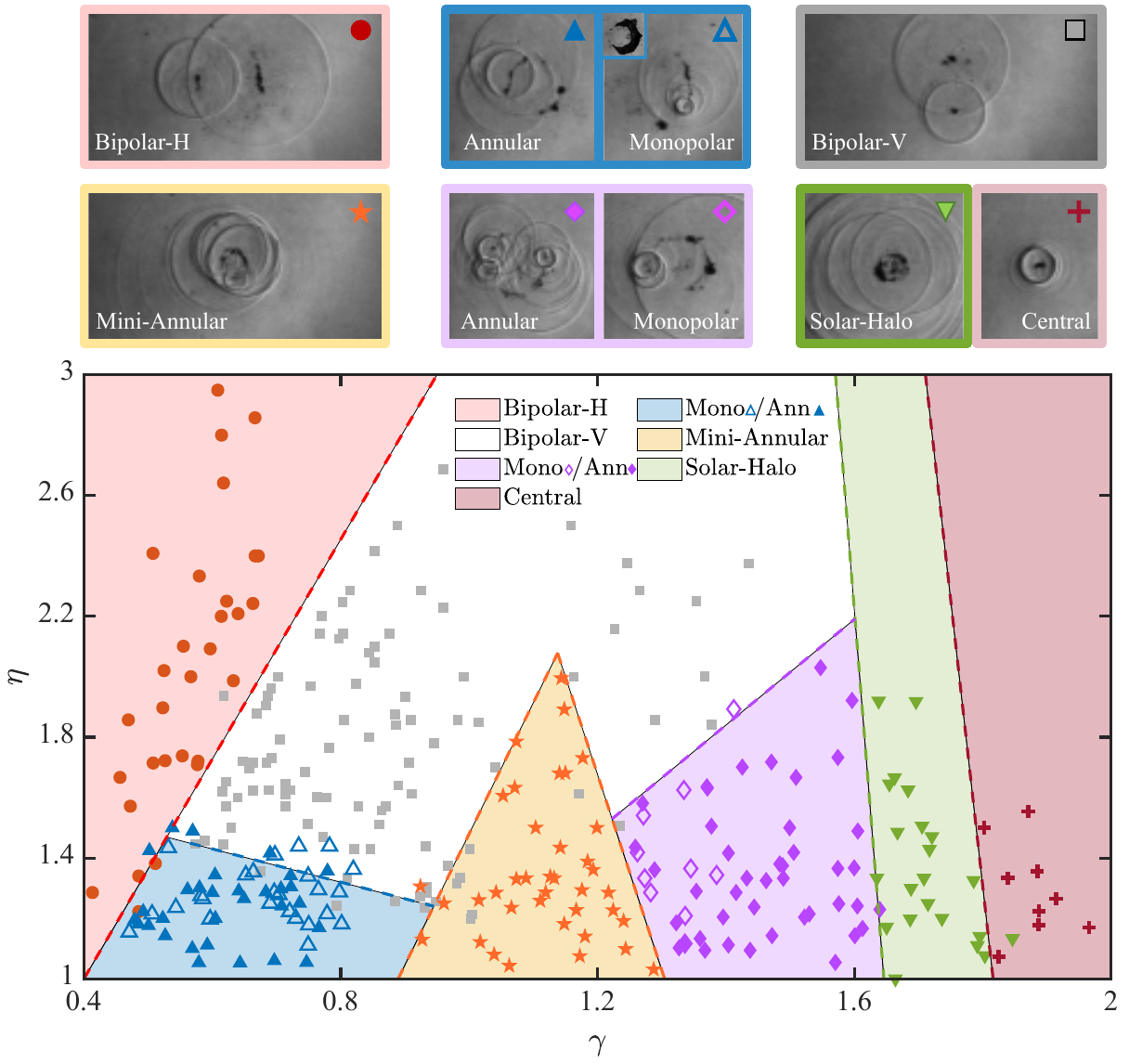}
    \caption{The regime map of six bubble collapse modes in the second cycle within the $\gamma$-$\eta$ space: Bipolar-H, Bipolar-V, Monopolar, Annular, Mini-Annular, Soloar-Halo, and Central. 
    The parameter space is divided into 7 regions, whose typical examples are shown above the regime map. 
    }
    \label{fig: eta-gamma}
\end{figure}

\subsection{The influence of initial nonsphericity on bubble collapse}\label{4.1}%%

The results presented earlier, as shown in figure \ref{fig: Rs}, have indicated that the collapse mode of cavitation bubbles is not exclusively influenced by the stand-off distance. 
Figure \ref{fig:eta-shock-wave} compares the collapse modes and shockwave characteristics of bubbles with different initial aspect ratios, $\eta$, at similar distance parameters in the range of $\gamma$ from 0.6 to 0.7: 
i) at $\eta=1.11$, bubbles with high initial sphericity collapse in the Annular mode; 
ii) at $\eta=1.41$, the increased asymmetry in the initial shape leads to the symmetry-breaking instabilities, causing the bubbles to collapse in the Monopolar mode; 
iii) at $\eta=1.81$, bubbles with a larger aspect ratio collapse in the Bipolar-V mode, with two polar regions located on either side of the laser beam; iv) at $\eta=2.61$, bubbles with an extremely elongated shape initially collapse in the Bipolar-H mode, with two polar regions distributed axially along the laser beam. The corresponding animations are provided in the supplementary movie \textcolor{red}{2} for the convenience of readers. 

Furthermore, we conducted over 300 experiments within the parameter space of $\gamma=0.4\sim2.0$ and $\eta=1.0\sim3.0$.
Based on the experimental observations, we plotted the regime map of bubble collapse modes in figure \ref{fig: eta-gamma}.
At $\gamma<0.9$, the predominant collapse modes are Bipolar-H, Bipolar-V, Monopolar, and Annular, consistent with the findings in figure \ref{fig: Rs}.
In most cases, the Monopolar and Annular modes exhibit varying degrees of randomness in their occurrence, coexisting within a trapezoidal region of the parameter space. However, when the initial bubble shape is nearly spherical ($\eta\le 1.1$), the collapse exclusively manifests as the Annular mode. The Bipolar-H is observed only at small stand-off distances ($\gamma<0.7$), with the corresponding critical $\eta$ increasing as $\gamma$ becomes larger. The transition mechanism between the Bipolar-H and Bipolar-V modes is an intriguing topic that will be further elaborated in \S\ref{4.2}. For $\gamma$ ranging from 0.9 to 1.2, the Mini-Annular mode dominates within a parameter region delineated by a yellow triangle, with its dynamic mechanism thoroughly analyzed in \S\ref{sec:Cavitation erosion mechanism}. 
As $\gamma$ increases further, Mono/Ann modes re-emerge in a quadrilateral regime. 
The Bipolar-V modes are extensively distributed within $0.5<\gamma<1.5$, occupying the largest area on the regime map. 
This indicates that the initial non-sphericity of the bubble has a profound and extensive influence on bubble collapse during the second cycle. The Solar-Halo and Central modes, which dominate at larger stand-off distances, are less sensitive to variations in $\eta$ due to the gradual attenuation of the initial asymmetry during bubble migration toward the boundary.

Overall, the initial shape of the bubble directly influences the distribution of energy and the focusing direction during collapse, particularly for $\gamma<1.0$. 
In natural cavitation, bubbles may also exhibit non-spherical shapes upon nucleation. When exposed to vortex flow or non-uniform pressure gradients, bubbles experience compression in certain directions, resulting in deformation into ellipsoidal forms \citep{dabiri2010interaction, dular2019high, hsiao2020dynamics}.
With increasing $\eta$, the collapse undergoes a transition from axially symmetric behavior to circumferentially asymmetric behavior.

\subsection{The mechanism for circumferential asymmetrical bubble collapse}\label{4.2}

\begin{figure}
    \centering
    \includegraphics[width=1\linewidth]{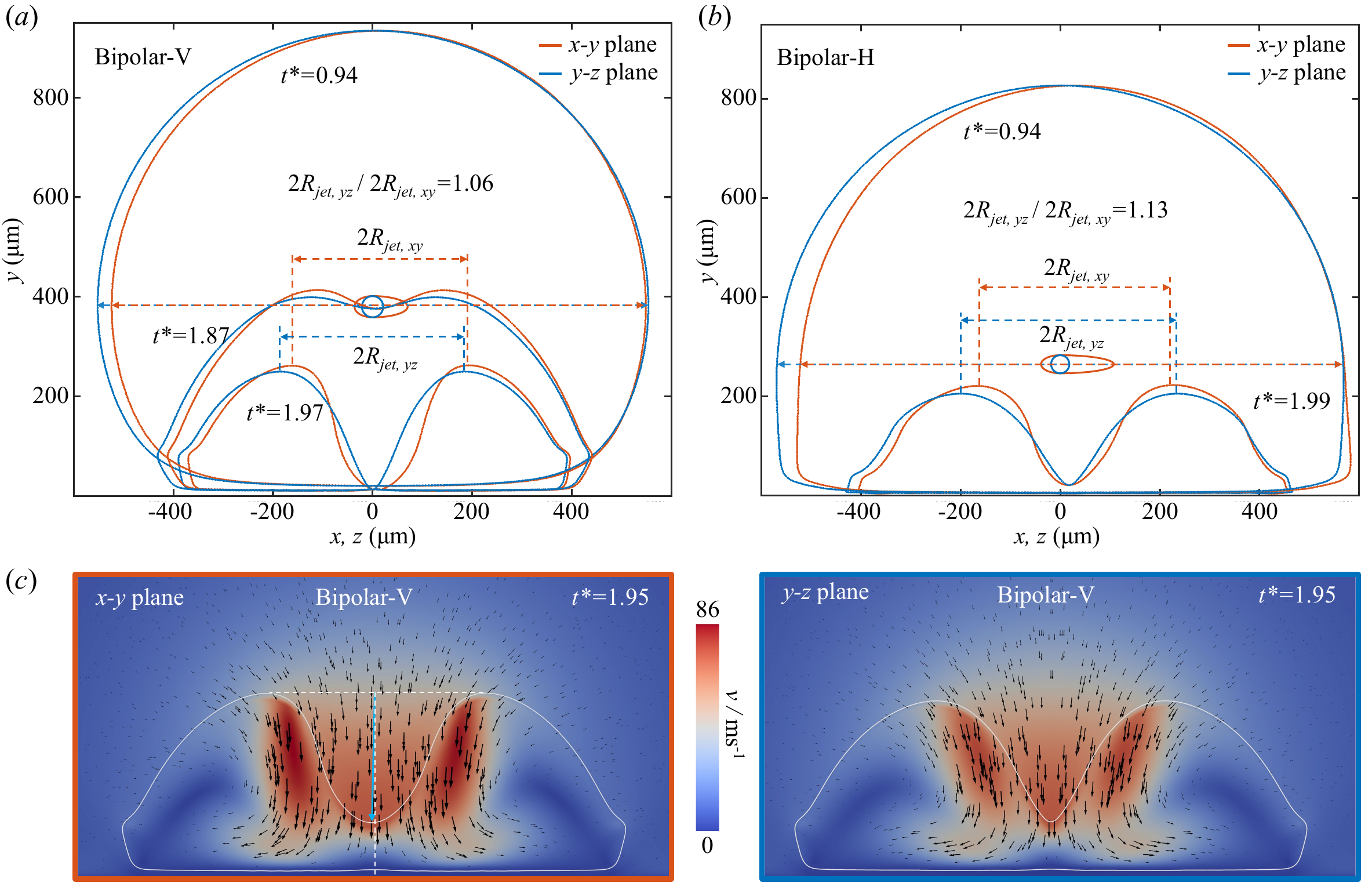}
    \caption{The evolution of an initially non-spherical bubble collapse near a rigid wall during its first cycle. The instantaneous bubble interface contours sliced from the $x$-$y$ plane and the $y-z$ plane for the bubble ultimately collapsed in  ($a$) the Bipolar-V mode ($\gamma=0.72$, $\eta=1.8$, $\xi=1.9$) and ($b$) the Bipolar-H mode ($\gamma=0.50$, $\eta=2.2$, $\xi=2.8$). $(c)$ A comparison of the sliced velocity fields for the Bipolar-V collapse mode at $t^* = 1.95$. The white curve showing the numerical bubble iso-surface at $\alpha=0.5$, and the black arrows indicate the velocity vectors.
    }
    \label{fig: first}
\end{figure}

Three-dimensional numerical simulations were conducted to reveal the mechanism behind circumferentially asymmetric collapse of the bubble, especially the Bipolar mode, triggered by initially non-spherical bubbles. Figures \ref{fig: first}$(a)$ and $(b)$ illustrate the time evolution of the contours of the bubbles' interfaces and flow fields during the first oscillation cycle. The bubbles originated from a ``teardrop-shaped'' form at nucleation ($t^*=0$) and collapsed in the Bipolar-V and Bipolar-H modes during the second cycle, respectively. 

\begin{figure}
    \centering
    \includegraphics[width=1\linewidth]{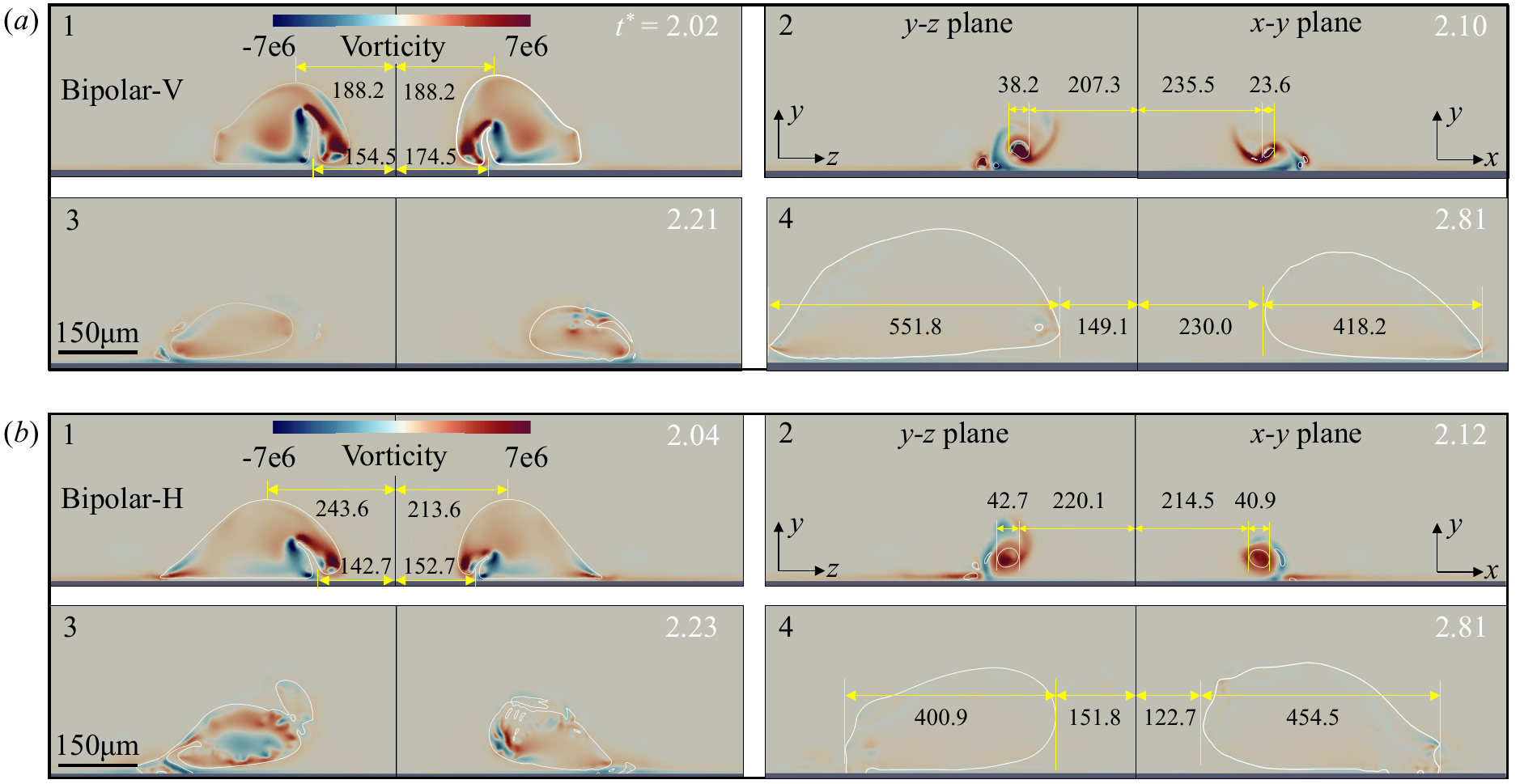}
    \caption{Evolution of the vorticity field of the initially non-spherical bubble after jet impact on the wall. ($a$) The Bipolar-V mode: $\gamma=0.72$, $\eta=1.8$, $\xi=1.9$; ($b$) the Bipolar-H mode: $\gamma=0.50$, $\eta=2.2$, $\xi=2.8$. The white curve showing the numerical bubble iso-surface at $\alpha=0.5$.}
    \label{fig: vorticity}
\end{figure}

\begin{figure}
    \centering
    \includegraphics[width=1\linewidth]{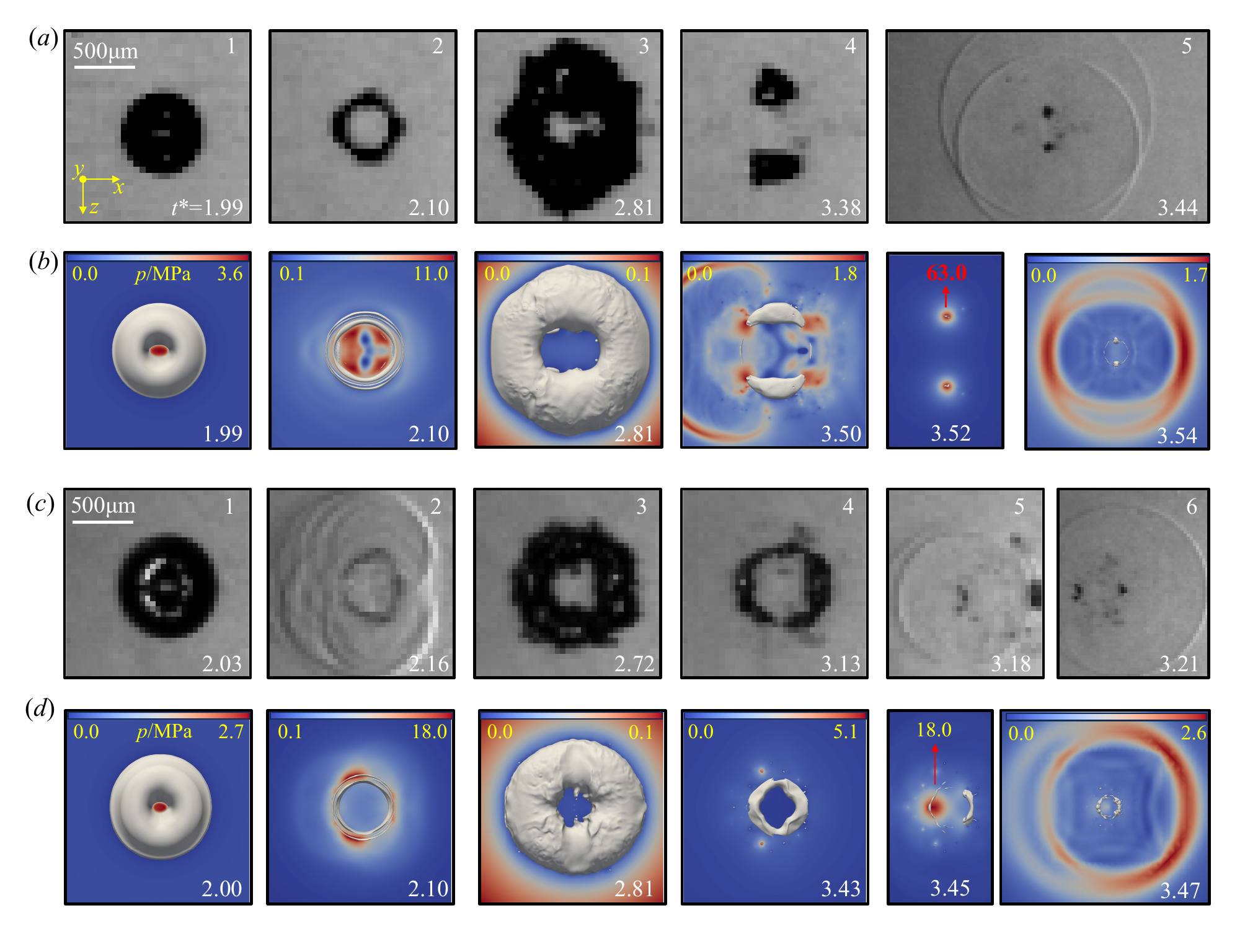}
    \caption{Comparison between experimental observations and numerical simulation results. Bipolar-V: $(a)$ \& $(b)$ $\gamma = 0.72$, $\eta = 1.8$, $\xi = 1.9$, $R_{\max} = 0.53$ mm. Bipolar-H: $(c)$ $\gamma = 0.61$, $\eta = 2.61$, $\xi = 2.3$, $R_{\max} = 0.59$ mm. $(d)$: $\gamma = 0.50$, $\eta = 2.2$, $\xi = 2.8$, $R_{\max} = 0.53$ mm. Contour map represents the pressure on the wall surface.}
    \label{fig: exp-num-compare}
\end{figure}

In the Bipolar-V mode, during the initial stage of bubble nucleation, the bubble exhibits a larger profile in the $x$-$y$ plane, requiring the displacement of more surrounding water as it expands. As expected, the greater added mass results in increased inertial resistance, leading to a slightly smaller expansion. When the bubble volume reaches its maximum, its maximum width in the $x$-$y$ plane is approximately 97\% of that in the $y$-$z$ plane, leading to a larger curvature at the bubble apex ($t^{*}=0.94$). 
Based on the Rayleigh equation \citep{duclaux2007dynamics}, bubble cross-sections with smaller apex curvature radii generate greater jet momentum during the collapse process. Since the jet maintains a consistent axial velocity across different planes (see figure \ref{fig: first}$c$), the jet in the $x$-$y$ plane, expands more in the transverse direction to increase its volume. In contrast, the jet profile in the $y$-$z$ plane is relatively narrower. This difference explains the formation mechanism of the elliptical jet tip observed in figure \ref{fig:side view}$(a)$. 
Additionally, the ``tear-drop shaped'' bubble exhibits asymmetry along its major axis ($\xi=1.9$), causing a deflection angle between the jet axis and the wall normal, as indicated by the blue arrows in figure \ref{fig: first}$(c)$. A similar phenomenon was observed in the fourth frame of the experiment in figure \ref{fig:bubble near wall exp}$(a)$, where the jet typically deflects towards the laser incidence direction. 
As shown in figure \ref{fig: vorticity}$(a)$, the elliptical jet tip leads to asymmetric splashing with respect to the wall center. The upward propagation speed of the annular splashing jet in the \(x\)-\(y\) plane is slower compared to that in the \(y\)-\(z\) plane, as shown in frame 1. The impact of the splashing jet on the upper surface of the toroidal bubble results in fragmentation of the vortex ring, where the stronger splashing leads to a relatively larger gas volume being retained within the primary vortex ring, closer to the central axis (see frame 2).
Nevertheless, the daughter bubbles coalesce into a larger one as they expand during the second cycle (frame 3). When the bubble reaches its maximum size (frame 4), it is observed that the bubble profile in the \(y\)-\(z\) plane is larger than that in the \(x\)-\(y\) plane. Consequently, the toroidal bubble collapses faster and pinches off in the \(x\)-\(y\) plane. Further, as shown in the 3D simulation and experimental comparison in figures \ref{fig: exp-num-compare}$(a)$ and $(b)$, the toroidal bubble exhibits smaller radii of curvature and volume at its left and right ends than at the top and bottom during expansion in the second cycle (see figure \ref{fig: exp-num-compare}($a3$)). During the subsequent collapse process, the toroidal bubble pinches off into an upper and a lower daughter bubble (see figure \ref{fig: exp-num-compare}$(a4-a5)$). 
From the wall pressure contour maps, it is evident that the water hammer pressure load induced by the jet is only 3.6 MPa, whereas the wall pressure during the first collapse of the toroidal bubble reaches 11 MPa. This indicates that the load during the initial collapse phase of the bubble is relatively small, insufficient to trigger the severe localized erosion phenomena depicted in figure \ref{fig:cavitation erosion}. However, as the bubble enters the second cycle, the ``lip-shaped'' bubble undergoes asymmetric collapse along the circumferential direction, leading to wavefront superposition effects (frame 4). This superposition of wavefronts intensifies the collapse strength at both the upper and lower ends of the bubble, inducing wall pressures up to 63 MPa. This pressure far exceeds the yield strength of pure aluminum, sufficient to cause plastic deformation of the material.

In the Bipolar-H mode, as shown in figure \ref{fig: first}$(b)$, an increase in \(\eta\) enhances the curvature difference between the two planes during bubble oscillation, similarly resulting in an elliptical jet tip. The jet in the \(y\)-\(z\) plane exhibits a significantly wider top radius than in the \(x\)-\(y\) plane, which limits the volume of the primary vortex ring near the central axis during the collapse phase (figure \ref{fig: vorticity}$b$, frame 1). When the bubble collapses to its minimum volume, the vortex ring volumes in both planes become nearly identical (frame 2). Note that the position of the cross-section centroid of the bubble in the \(y\)-\(z\) plane is larger compared to the \(x\)-\(y\) plane. As a result, as shown in figures \ref{fig: exp-num-compare}$(c)$ and $(d)$, the top and bottom ends of the toroidal bubble have a smaller oscillation period and collapse earlier in the \(y\)-\(z\) plane during the second cycle \citep{li2020dynamics}. This is responsible for the pinch-off of the toroidal bubble into a left and a right daughter bubbles. 
Additionally, as observed in the second frame of figure \ref{fig: exp-num-compare}$(c)$, the left end of the toroidal bubble collapses first (possibly due to a larger jet deflection angle), releasing a shockwave, which then converges circumferentially towards the right endpoint. This process ultimately forms a shockwave interference front, similar to the Monopolar \uppercase\expandafter{\romannumeral1} mode shown in figure \ref{fig: shock-wave-diagram}. The time phase difference in the collapse of the toroidal bubble on the left and right sides also carries over to the second cycle. It is worth noting that the wall pressure caused by the Bipolar-H mode is only 18 MPa, which is significantly lower than that of the Bipolar-V mode.

In summary, the asymmetry of the initial bubble shape results in differences in added mass and inertial resistance during bubble expansion in different planes, thereby forming a jet with an elliptical tip. The variations in the cross-sectional area of the toroidal bubble and the position of the cross-section centroid control the transition between the Bipolar-H and Bipolar-V collapse modes.

\subsection{The impact pressure induced by circumferential asymmetrical bubble collapse}\label{4.3}

\begin{figure}
    \centering
    \includegraphics[width=1\linewidth]{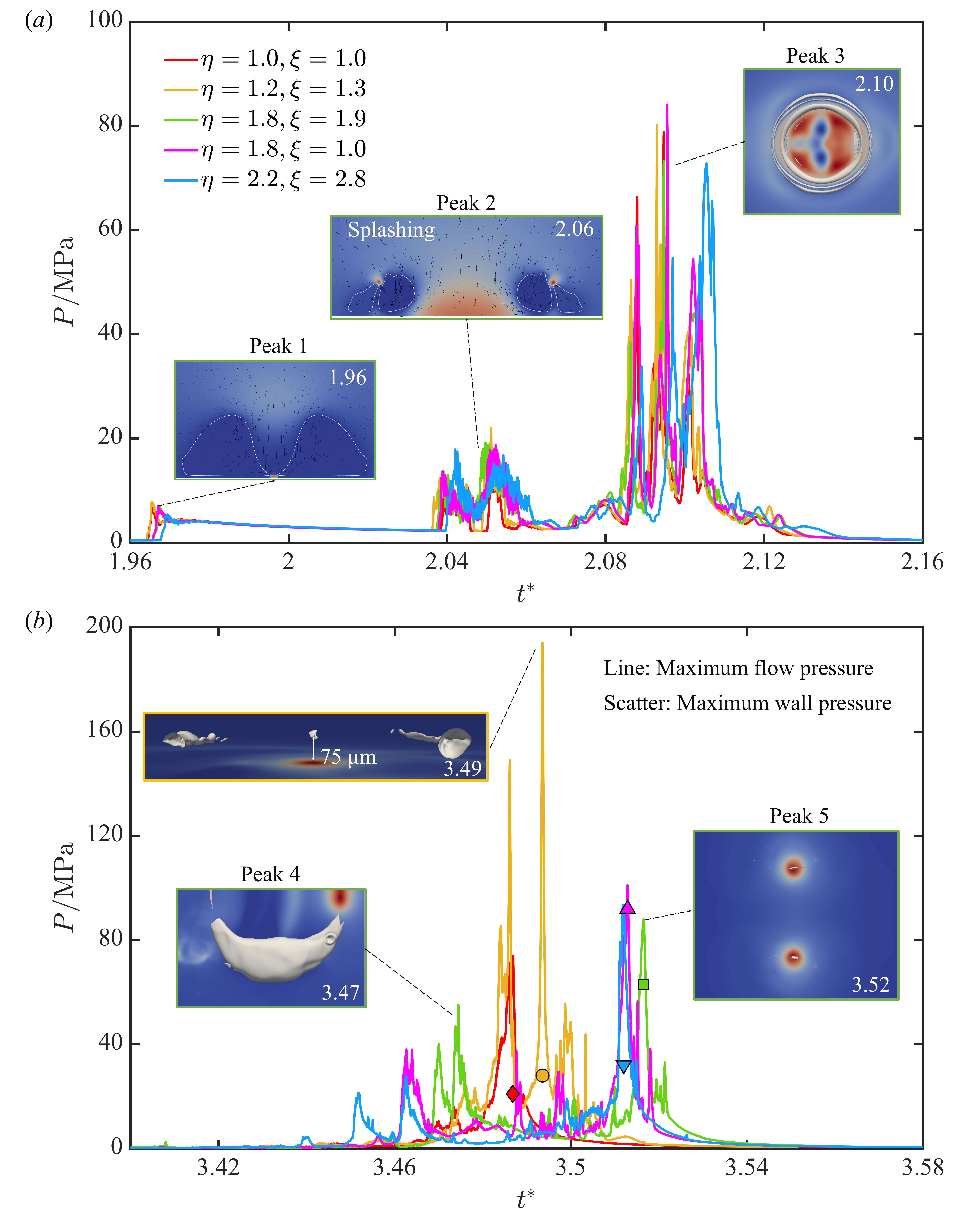}
    \caption{Comparison of the temporal evolution of the pressure induced by bubbles ($R_{\max} \approx 0.53$ mm, $\gamma \approx 0.72$) with varying initial shapes: $(a)$ the first collapse; $(b)$ the second collapse. A counterpart animation is provided in the supplementary movie \textcolor{red}{2}.}
    \label{fig: P-A-compare}
\end{figure}

\begin{figure}
    \centering
    \includegraphics[width=1\linewidth]{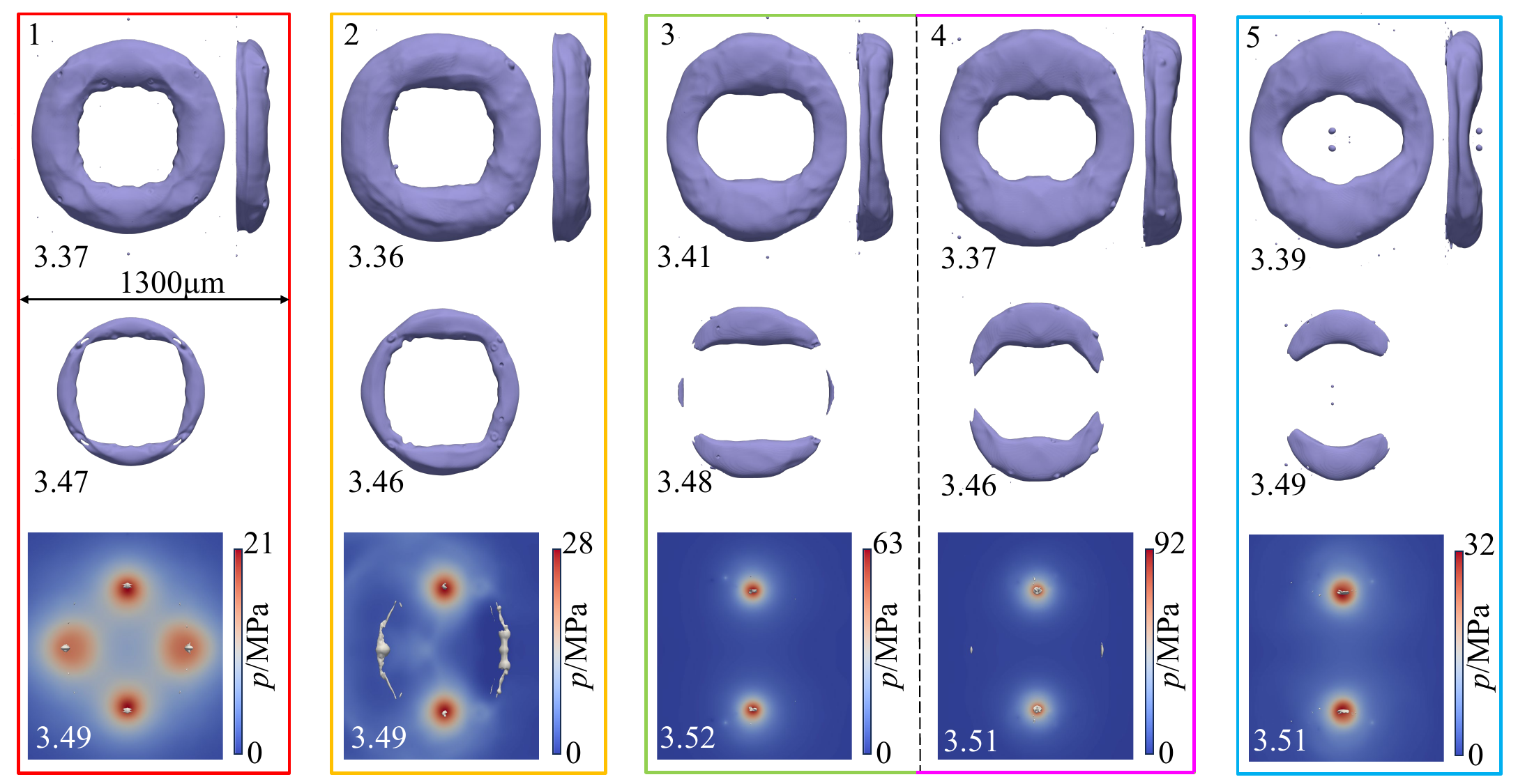}
    \caption{The 3D numerical results ($R_{\max} \approx 0.53$ mm, $\gamma \approx 0.72$) demonstrate the influence of the initial bubble shape on the second collapse mode. The characteristic parameters from left to right cases are: i) $\eta = 1.0$, $\xi = 1.0$; ii) $\eta = 1.2$, $\xi = 1.3$; iii) $\eta = 1.8$, $\xi = 1.9$; iv) $\eta = 1.8$, $\xi = 1.0$; v) $\eta = 2.2$, $\xi = 2.8$. }
    \label{fig: annual-compare}
\end{figure}

To determine whether the localized high-pressure phenomenon discussed in \S\ref{4.2} occurs only when the initial bubble is non-spherical, we conducted five comparative numerical simulations with different initial bubble shapes under the same stand-off distance. 
The results indicate that variations in the initial shape have little impact on the bubble oscillation period, and detailed analysis is provided in \ref{appendix c}. 
We then compared the time evolution of the pressure field during the first and second collapse stages, as shown in figure \ref{fig: P-A-compare}. 

For the first bubble collapse, the pressure curve exhibits a three-peak feature, corresponding to three significant physical phenomena: jet impact on the wall, liquid ``splashing'', and the toroidal bubble collapse. 
First, as indicated by Peak 1, the jet penetrates the bubble's lower surface and impacts the wall at a velocity of about 70 m/s, forming a localized high-pressure region. 
Notably, a thin liquid film of about 10.6 $\upmu$m exists between the bubble's lower surface and the wall. 
According to \cite{reuter2019high}, using Total Internal Reflection (TIR) shadowgraphy, it was observed that for $\gamma < 1.0$, the liquid film thickness during bubble collapse ranges from 5 to 40 $\upmu$m, suggesting that the jet impact does not directly reach the wall, thereby mitigating the wall pressure to some extent. 
Through numerical simulations, we found that the maximum jet impact pressure is attenuated by approximately 50\% when transmitted to the center of the wall through a water film with a thickness of 10.6 $\upmu$m. 
A comparison of the curves for different $\eta$ values shows that as the bubble deviates from a spherical shape, the wall pressure induced by the jet decreases as $\eta$ increases. 
After the jet penetrates the bubble, it spreads radially along the wall, forming an annular liquid film in a direction different from the jet. This phenomenon, often referred to as the ``splashing effect'' or ``secondary jet'' by researchers \citep{tong1999role, brujan2002final}, generates significant pressure (Peak 2) when the rapidly expanding splashing edge collides with the upper surface of the collapsing bubble. 
However, since this phenomenon occurs far above the wall, it induces high local pressure but significantly attenuates before reaching the wall. 
When the bubble collapses to its minimum volume, Peak 3 is formed, with its peak value approximately ten times that of Peak 1. However, since the collapse of the toroidal bubble is not simultaneous at all locations, multiple pressure peaks are generated at different positions. 
Notably, when $\eta = 1.8$, the ellipsoidal initial shape ($\xi = 1.0$) induces a higher flow field pressure compared to the case with $\xi = 1.9$.

During the second collapse of the bubbles, the peak pressure of the flow field induced by bubble collapse was higher than that of the first collapse, consistent with the analysis of the cavitation erosion mechanism in \S\ref{3.4}. Furthermore, except for $\eta = 1.0$, the flow field pressure during the second collapse exhibited a distinct double-peak feature.
Figure \ref{fig: annual-compare} presents three typical frames from the second collapse cycle: the top row shows the morphological characteristics of the toroidal bubbles during the contraction process (including front and side views); the middle row depicts the tearing state of the bubbles in the late collapse stage; and the bottom row illustrates the moment when the bubbles collapse to their minimum volume.

i) At $\eta=1.0$, the toroidal bubble exhibited a uniform thickness and contracted at similar speeds across its surface. The curvature within the ring showed minimal variation, with only slightly elevated local curvatures at the four corners. The bubble tore into four sections and collapsed simultaneously at $t^{*} = 3.49$, exhibiting a single-peak characteristic. This process generated a significant pressure in the flow field, but the wall pressure reached only 21 MPa. 

ii) For $\eta=1.2$, the bubble ring also maintained a relatively uniform thickness, with slightly increased curvature at the four corners. The left and right sides collapsed earlier than the top and bottom sides. Assuming the shock wave propagates at the speed of sound (1500 m/s) towards the top and bottom residual bubbles, over a distance of approximately 410 $\upmu$m, the estimated time is 0.27 $\upmu$s. The time interval between Peak 4 and Peak 5 is approximately 0.4 $\upmu$s. This indicates that the shock waves from the left and right sides enhanced the collapse of the top and bottom bubbles, resulting in a maximum pressure of 192 MPa in the flow field. However, the collapse occurred at a distance of approximately 75 $\upmu$m from the wall (figure \ref{fig: P-A-compare}), leading to a wall pressure peak of only 28 MPa, significantly lower than the flow field pressure peak.

iii) For $\eta=1.8$ and $\xi=1.9$, the curvature on the left and right sides of the bubble increased significantly during the secondary oscillation. The side view showed that the central thickness of the bubble ring was approximately 62\% of the thickness at the top and bottom. The thinner and narrower regions collapsed and disappeared first, producing a pressure peak (Peak 4) in the flow field, as shown in the subplot for $t^{*} = 3.47$ in figure \ref{fig: P-A-compare}$(b)$. The final lip-shaped bubble collapse generated a localized wall pressure peak reaching 63 MPa (Peak 5).

iv) Notably, for $\eta=1.8$ and $\xi=1.0$, the overall collapse pattern of the bubble was similar to the case of $\xi=1.9$, indicating that axial asymmetry had a limited influence on the overall collapse behavior. However, the wall pressure peak caused by $\xi=1.0$ reached 92 MPa, nearly equal to the maximum pressure in the flow field and approximately 1.46 times that for $\xi=1.9$.

v) For $\eta=4.0$, the asymmetry between the top/bottom and left/right ends of the bubble became more pronounced, with the width approximately twice that of the sides and the thickness about 1.7 times greater. The Bipolar-V collapse mode was observed again. However, the wall pressure peak was significantly reduced, reaching only 32 MPa. This is also due to the attenuation effect of water on the shock wave load.

Overall, at moderate stand-off distances, the flow field pressure induced by the secondary collapse of the bubble is higher than that of the initial collapse. As $\eta$ increases, the circumferential asymmetry of bubble collapse becomes more pronounced. Additionally, under the same aspect ratio, elliptical initial bubbles generate higher wall pressure compared to teardrop-shaped bubbles, reaching up to three times that of spherical bubbles. 

Finally, we replicated an extreme case studied by \cite{rodriguez2022dynamics} (figure 2$a$ therein), with \(\gamma = 1.25\), and the peak pressure is approximately 400 MPa (the comparison results are not shown here). Our numerical results exhibit strong similarities in bubble collapse patterns and jetting behaviors. While the timing of the peak pressures between the two models varies by only 1\%, the amplitude of the peak pressure shows a deviation of about 40\%. This discrepancy may arise from the lower accuracy of the present finite volume method (FVM) in capturing the shockwave front compared to higher-order methods. Additionally, thermal effects were not considered in our model \citep{beig2018temperatures, zhou2020modelling, white2023pressure}, which could also contribute to the observed differences. Future work will focus on investigating the influence of thermodynamic effects and ambient temperature on cavitation erosion.

\section{Conclusions}\label{sec:Conclusion}

This study presents a comprehensive experimental and numerical investigation of the erosion of aluminum samples induced by laser-induced cavitation bubbles at moderate stand-off distances (\(\gamma = 0.4 \sim 2.2\)). We first classified and analyzed five typical erosion patterns, namely Bipolar, Monopolar, Annular, Solar-Halo, and Central. Quantitative analysis of the erosion damage and associated bubble collapse patterns during the first and second cycles of the bubble provides new insights into the mechanisms responsible for the significant erosion.

An image-based quantification method was established to assess cavitation erosion damage, revealing significant peaks in cavitation erosion damage volume at \(\gamma\approx0.6\) and \(1.4\). Using Gaussian multi-peak fitting, we quantified the radial erosion depth distribution and determined the radial position of the maximum erosion depth, typically occurring within \(R^{*} = R/R_{\text{max}} < 1\). By combining bubble pulsation and shock wave release characteristics, we demonstrated that erosion pits are primarily caused by shock waves generated during bubble collapse rather than jet impact. In particular, the secondary collapse of the toroidal bubble leads to converging wavefronts colliding head-on and shock waves superposing obliquely, which are the primary mechanisms for severe local cavitation in Bipolar and Monopolar modes. These modes are attributed to the circumferentially asymmetrical collapse of the bubble. Notably, we found that erosion pits do not always align with the toroidal bubble radius during the second collapse process, especially in the Mini-Annular collapse mode.

We demonstrated that the initial shape of the bubble has a significant influence on the collapse mode and the impact pressure. Experimental and numerical results have confirmed that as the aspect ratio $\eta$ increases, the asymmetric characteristics of bubble collapse become more pronounced, and a new phenomenon called Bipolar-H is observed at large $\eta$. The variations in the cross-sectional area of the toroidal bubble and the position of the cross-section centroid control the transition between the Bipolar-H and Bipolar-V collapse modes. A regime map of bubble collapse modes in the $\gamma-\eta$ parameter space was plotted based on extensive experimental results. 

Lastly, our 3D numerical model reveals that the difference in inertial resistance during bubble expansion across different planes leads to the formation of an elliptical jet tip, accurately reproducing the collapse phenomenon of the 'lip-like' bubble in the Bipolar mode. Moreover, it was confirmed that initially elliptical bubbles undergoing the head-on collision of convergent wavefronts generate intense wall impact loads. The peak load can reach up to 92 MPa, exceeding that of bubbles with an initial spherical shape by more than threefold under similar conditions.

\appendix
\renewcommand{\thesection}{\textbf{Appendix \Alph{section}}}
\section{Image-based recognition method for cavitation erosion pits}\label{appendix A}

First, the erosion pattern was recognized as a binarized figure using ImageJ, and micro pits taking only one pixel were filtered out as noise.
Then, we developed MATLAB codes to identify the closed regions, where the voids were filled to form single-connected patterns. 
Each single-connected pattern was considered an independent pit, and the radius ($R_{\rm pit}$) and position of the equivalent circular pit, with the same area and centroid, could be determined (figure \ref{fig: image-processing}$b$), similar to the Pit-Count method proposed by 
\cite{dular2004relationship}.
The data processing code used in this study has been made publicly available on GitHub at https://github.com/Zhesheng111/CavEro.

\begin{figure}
    \centering
    \includegraphics[width=1\linewidth]{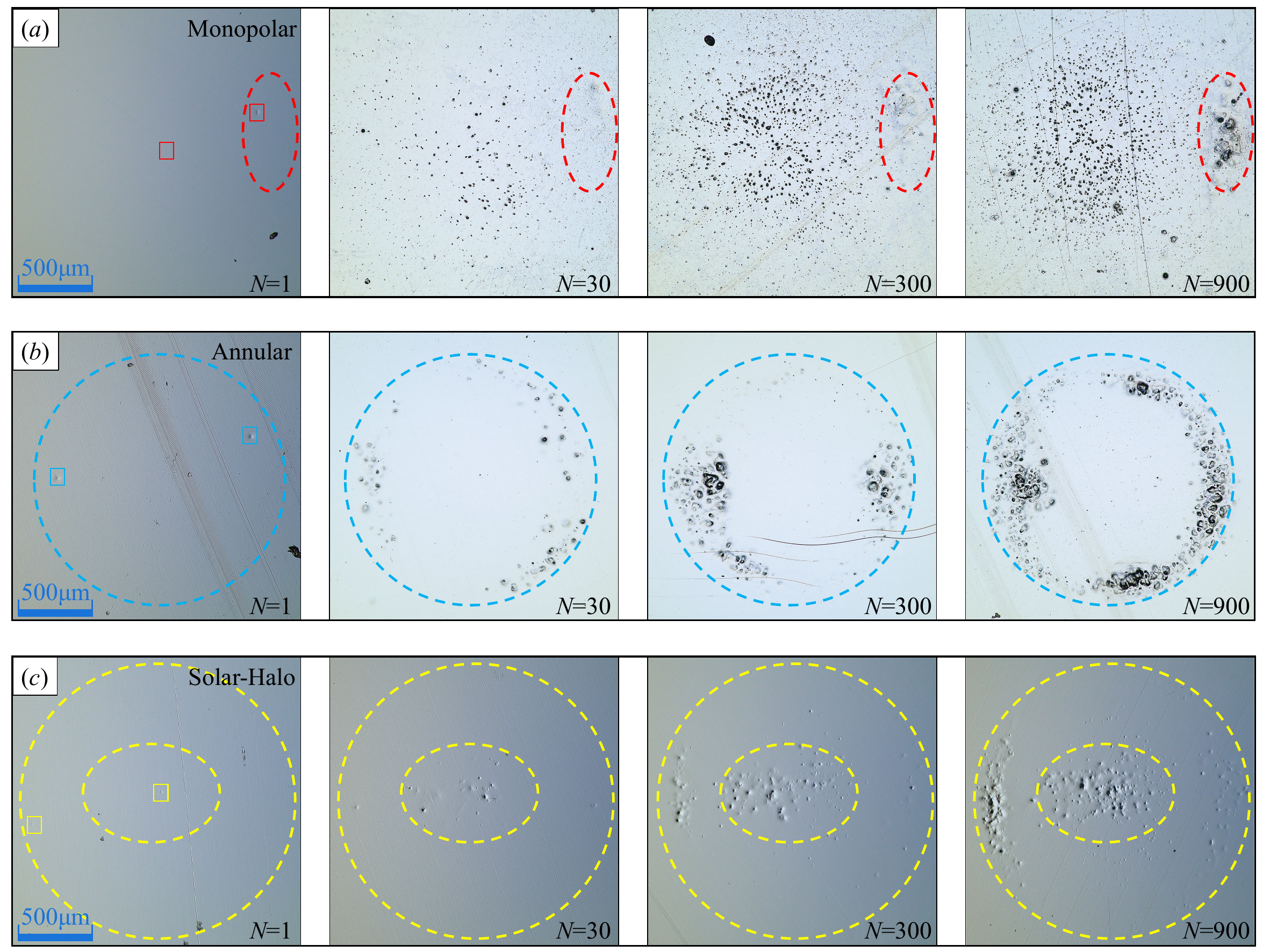} 
    \caption{Illustration of the effect of bubble number N on cavitation erosion distribution characteristics: $(a)$ Monopolar, $R_{\max}$ = 1.12 mm, $\gamma = 0.75$; $(b)$ Annular, $R_{\max}$ = 1.04 mm, $\gamma = 1.49$; (c) Solar-Halo, $R_{\max}$ = 1.00 mm, $\gamma = 1.77$. The darker images were taken under oblique illumination, while the brighter ones were captured under bright-field lighting. 
    }
    \label{fig:bubble-N}
\end{figure}

\renewcommand{\thesection}{\textbf{Appendix \Alph{section}}}
\section{Impact of bubble number on cavitation erosion patterns}\label{appendix B}

We selected three representative stand-off distances and systematically compared the cavitation erosion patterns induced by a single bubble ($N = 1$) and multiple repeated bubbles ($N = 30, 300, 900$), as shown in figure \ref{fig:bubble-N}. The brighter panels were captured under bright-field illumination, while the darker ones were taken under oblique illumination to reveal shallow indentations. In each frame, the dashed region highlights the most prominent erosion feature, while the boxes in the first frame mark the pits caused by a single bubble. 
The severity of cavitation erosion increases with the number of bubbles. Pre-existing damage areas expand further due to the accumulation and overlap of pits, leading to more pronounced erosion features. Remarkably, single-bubble pit locations align with erosion regions in multi-bubble cases. For example, when $\gamma = 0.75$, the single bubble generated two pits—one at the center of the sample and the other within the pit aggregation region characteristic of the Monopolar pattern. When $\gamma = 1.49$, the pits were distributed along the periphery, with no damage observed at the center, even at $N = 900$. For the Solar-Halo pattern, at $N = 1$, shallow indentations appeared at both the center and periphery, suggesting that the central pits observed at $N$ $\neq$ 1 result from secondary bubble collapse rather than shock wave-induced residual bubble collapse. 
The results show that as the number of bubbles increases, the erosion pits exhibit a statistically consistent distribution. Even if the sample surface undergoes deformation and damage, the collapse mode of the bubbles remains unaffected.

\renewcommand{\thesection}{\textbf{Appendix \Alph{section}}}
\section{Effect of initial bubble shape on bubble oscillation}\label{appendix c} 

\begin{figure}
    \centering
    \includegraphics[width=0.90\linewidth]{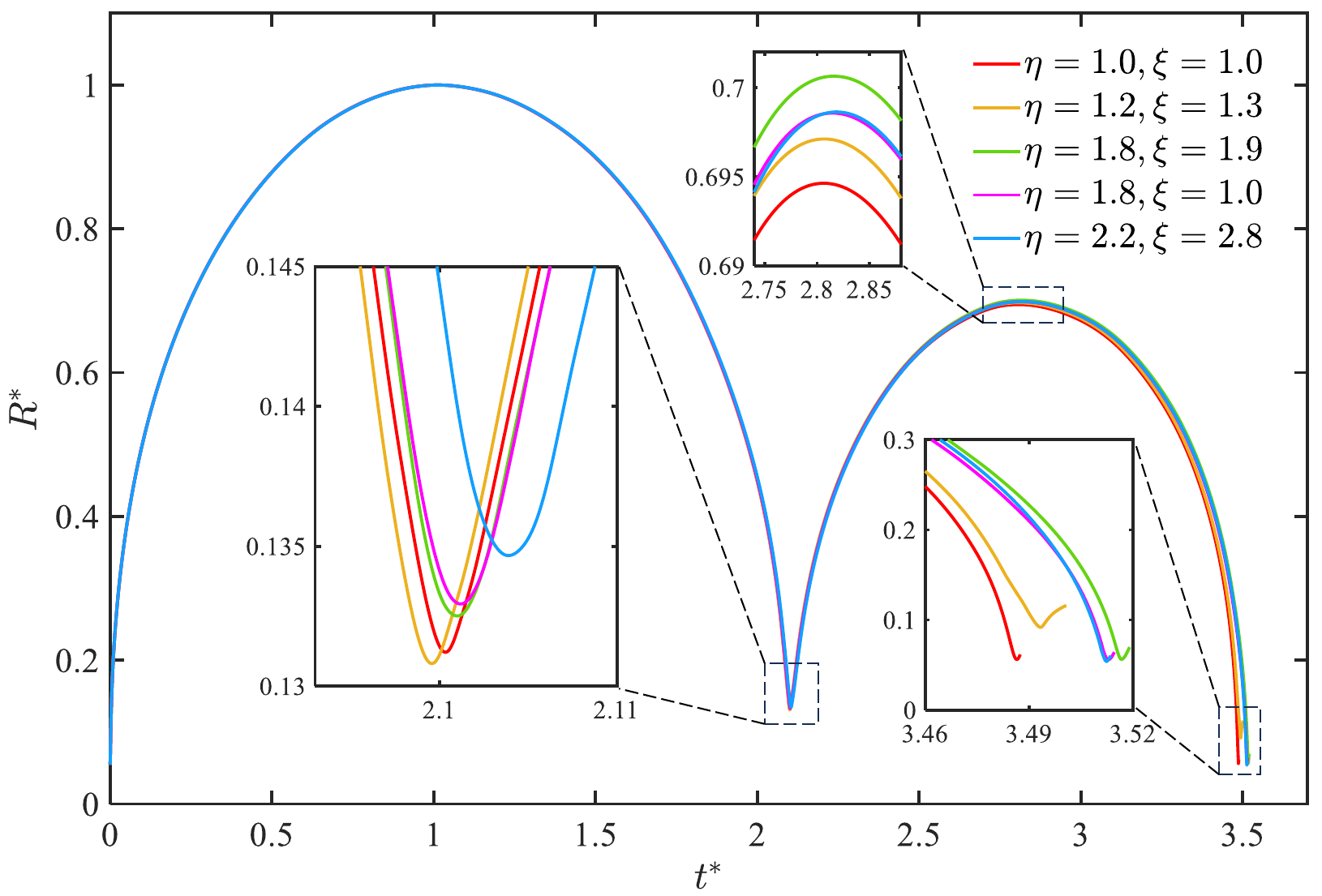}
    \caption{Comparison of the time-history curves of the bubble pulsation radius near a wall for different initial bubble shapes. The maximum bubble radius and stand-off distance parameters are similar in each case ($R_{\max} \approx 0.53$ mm, $\gamma \approx 0.72$). The characteristic parameters are $\eta = 1.0$, $\xi = 1.0$; $\eta = 1.2$, $\xi = 1.3$; $\eta = 1.8$, $\xi = 1.9$; $\eta = 1.8$, $\xi = 1.0$; and $\eta = 2.2$, $\xi = 2.8$. }
    \label{fig: radius-compare}
\end{figure}

Figure \ref{fig: radius-compare} presents the temporal evolution of the bubble radius, non-dimensionalized by the equivalent spherical radius corresponding to the bubble's maximum volume. 
The figure indicates that the initial bubble shape has negligible influence on the equivalent radius and oscillation period of the bubble. 
Notably, the magnified subplots reveal that when the initial bubble deviates from a spherical shape ($\eta > 1.0$), the minimum volume during the first collapse increases with increasing $\eta$, and the oscillation period becomes slightly longer. In addition, the bubble with a spherical initial shape exhibits the smallest rebound volume during the second cycle, indicating the greatest energy loss, which may lead to a reduction in collapse pressure. 
Furthermore, for the case of $\eta = 1.8$, the effect of the axial asymmetry parameter $\xi$ was studied. The results show that different $\xi$ values have almost no effect on the first bubble pulsation. However, the second pulsation period is slightly shorter when $\xi = 1.0$.

\section*{Acknowledgement}
The authors would like to acknowledge the support from National Natural Science Foundation of China (No.12332014, No.12372268), National Major Basic Research Project (No. KY61700230031), Natural Science Foundation of Heilongjiang Province (No.YQ2022A004) and Key R\&D Program Project of Heilongjiang Province (JD24A002).  
Deep gratitude is extended to Prof. Qingyun Zeng (HEU) and Prof. Yunlong Liu (HEU) for providing constructive suggestions on the numerical simulations.
Thanks also to Prof. Yiwei Wang (IMCAS) and Dr. Jingzhu Wang (IMCAS) for their assistance with experimental configurations.

\section*{Declaration of Interests} 
The authors report no conflict of interest.

\bibliographystyle{jfm}
% Note the spaces between the initials
\bibliography{SBL2}
\end{document}